\documentstyle[12pt,aaspp4,flushrt,psfig]{article}
\begin{document}

%

\title{RXTE MEASUREMENT OF THE DIFFUSE X-RAY EMISSION \\ 
FROM THE GALACTIC RIDGE: IMPLICATIONS FOR THE \\
ENERGETICS OF THE INTERSTELLAR MEDIUM}

\author{
Azita Valinia\altaffilmark{1} and  
Francis E. Marshall\altaffilmark{2}}  

\affil{Laboratory for High Energy
Astrophysics\\
 Code 662, NASA/Goddard
Space Flight Center, Greenbelt, MD 20771} 

\altaffiltext{1}{NAS/NRC Research Associate;
valinia@milkyway.gsfc.nasa.gov}
\altaffiltext{2}{marshall@milkyway.gsfc.nasa.gov}

\begin{center}
Accepted for Publication in the Astrophysical Journal  
\end{center}

\begin{abstract}
The diffuse X-ray emission from the thin disk surrounding the Galactic
mid-plane (the so-called Galactic ridge) was measured 
with {\it RXTE} PCA in order to determine the spatial extent, 
spectral nature, and origin of the emission. 
Spatial examination of the diffuse emission in the central
$30^\circ$ of the plane in Galactic longitude reveals the
presence of two components: a thin disk of full width $ \lesssim 0^\circ
\!.5$ centered roughly on the Galactic mid-plane, and a broad
component which can be approximated as a Gaussian distribution 
with FWHM of about $ 4^\circ$.
Assuming an average distance of 16~kpc to the edge of the
galaxy, a scale height of about $ 70$~pc and 500~pc
is derived for the thin 
and broad disk components, respectively. 
Spectral examination of the emission clearly reveals 
the presence of a hard power law tail above 10~keV and 
an emission line from He-like iron, 
indicating both thermal and possibly non-thermal
origins for the diffuse emission. 
The averaged spectrum from the ridge in the $3-35$~keV band can be 
modelled with a Raymond-Smith plasma component of temperature $\sim 2-3$~keV
and a power law component of photon index $\sim 1.8$.  
Based on this finding, we argue that the temperature of the hot phase
of the Interstellar Medium (ISM) is less than the previously reported 
values of $5-15$~keV. 

Motivated by the similarities between the characteristics of the 
thermal component of the Galactic ridge emission in our model and
the thermal emission from supernova remnants (SNRs), we discuss  
the origin of the thermal emission in terms of a population of  
SNRs residing in the 
Galactic disk. We find that a SN explosion rate of less than 5 per century
is adequate to power the thermal emission from the ridge.  
The origin of the
emission in the hard X-ray band modelled by a power law
remains uncertain. 
Possible contributions from non-thermal bremsstrahlung 
of cosmic ray electrons and protons,  
inverse Compton scattering  of energetic electrons from ambient microwave,
infrared, 
and optical photons, non-thermal emission from SNRs, and emission from 
discrete X-ray sources are discussed. 
We speculate that bremsstrahlung of accelerated electrons and protons in SNR
sites can play a significant role in producing the hard tail of the
spectrum. Moreover, their collisional losses can play a major role in the
ionization of the ISM.

\end{abstract} 
 

\newpage
%
 
\section{INTRODUCTION}

A galaxy can be well described by an ecosystem. There is an
intimate relationship, much like a symbiosis, between the discrete
components of the galaxy such as stars, and its interstellar medium (ISM).
The ISM provides the foundation for the birth of a new generation of stars, 
while at the same time it is enriched by the remains of the older
generations and their byproducts during their life cycle. 
Hence, it is natural to expect that the understanding of formation
and evolution of galaxies is closely related to the understanding of their ISM. 

The ISM of the Milky Way has been found to have many components. Magnetic
fields and cosmic ray gas compose the relativistic fluid, while the
gaseous phase consists of both ionized and neutral components. 
The hot ionized medium observable in UV
and X-ray has a temperature above $\sim 10^5$~K and is composed of hot coronal
gas heated by supernova shocks. A good portion of the energy of the ISM
resides in this component. The warm ionized medium (e.g. HII, planetary
nebulae) is visible in ${\rm H\alpha}$, UV, and optical, and has a
temperature as high as $10^4$~K. The neutral atomic gas appears to have
both cold ($<100$~K; e.g. HI~clouds) and warm (100~K~$<T<$~8000~K) components.  
Molecular clouds compose the self-gravitating gaseous phase. Finally the
ISM (with the exception of its hot phase) is filled with dust, 
particles typically a few
tenth of microns in size and visible through their infrared radiation.   
Since the peak emission of each component arises in different wavebands,
multi-wavelength observations from radio to $\gamma$-rays are needed 
to probe the ISM. 

Diffuse X-ray emission from our galaxy is a powerful diagnostic of
the hot phase of the ISM. It is our purpose here to use this tool
to probe the processes that contribute to its energetics and dynamics.
The first detection of the X-ray emission from the Galactic disk was
achieved by the pioneering rocket experiment of Bleach et al. (1972). 
They detected excess emission associated with a narrow disk component 
of angular size $2^\circ-7^\circ$. Since then, 
X-ray emission from the Galactic plane and in particular the 
ridge (the narrow region
centered on the Galactic mid-plane extending approximately 
to $\pm 60^0$ in longitude and
$\pm 10^0$ in latitude) has been measured in the past with several 
satellites (e.g. {\sl HEAO-1} [2-50~keV]: Worrall et. al. 1982;
{\sl EXOSAT} [2-6~keV]: Warwick et al. 1985; {\sl Tenma} [2-11~keV]: 
Koyama et al.
1986; {\sl Ginga} [2-16~keV]: Yamasaki et al. 1997; 
{\sl ASCA} [0.5-10~keV]: Kaneda et al. 1997).  
The presence of the $6.7$~keV iron line in the spectrum  
discovered with {\sl Tenma} has motivated
the idea that part of the emission below 10~keV 
is due to a hot optically thin plasma of temperature
$5-15$~keV. Because of its high spectral resolution, 
{\sl ASCA} has provided the most accurate
measurement of the spectrum of the emission to date. 
The presence of Mg, Si, and Fe K-lines in the spectrum obtained by {\sl ASCA} 
suggests that at least part of the emission is of thermal origin.
Close examination of {\sl ASCA} data has also revealed that unresolved, 
discrete sources are not responsible for the bulk of the emission
(Yamauchi et al. 1996; Kaneda 1997). 
The most recent investigation of the diffuse emission from the Scutum arm 
region with {\sl ASCA} (Kaneda et al. 1997) has concluded that the emission below
10~keV has both soft ($kT\sim 0.8$~keV) and hard ($kT \sim 7$~keV) thermal 
components. If indeed the super hot gas ($\sim 7$~keV) exists   
in an extended form in the ISM, it is not clear how to explain its 
confinement to the Galactic disk since its temperature exceeds the gravitational
potential of the disk by at least an order of magnitude (Townes 1989). 
Unfortunately, the two-temperature model does not produce a good fit to the
data above 10~keV indicating the presence of additional component(s) at
higher energies. Indeed, a hard power law tail has been detected in the
hard X-ray/soft $\gamma$-ray band from observations of the ridge with
{\sl Ginga} and the balloon
experiment {\it Welcome-1} (Yamasaki et al. 1997), and {\sl OSSE} (Skibo et al.
1997). 

In this paper, we present the results from {\sl RXTE} 
measurement of the diffuse
X-ray emission from the Galactic ridge in the 3-35~keV
band. Observations in the hard X-ray/soft $\gamma$-ray band have
usually been complicated by the presence of
numerous variable discrete sources, and the fact             
that the instruments generally have large
fields of view and no imaging capabilities, or have imaging
capability but no diffuse emission sensitivity. The combination of
these factors makes the separation of
emission between diffuse and compact sources a difficult task.
The advantage of {\sl RXTE} over
previous missions is its small field of view  
($1^\circ$ FWHM) combined with its wide energy bandpass (2-60~keV for the PCA), 
allowing for the 
subtraction of the contribution of discrete sources from the diffuse
emission spectrum in the hard X-ray band. 
In addition to reporting on the detection of a hard power law tail in the
{\sl RXTE} data, we also offer an 
alternative interpretation for the origin of the emission below 10~keV 
(i.e. instead of a super hot plasma of temperature
$5-15$~keV). 
In agreement with previous studies, our results indicate that 
the emission is most likely of
diffuse origin as opposed to the superposition of discrete sources. 
However, we present a model in which the X-ray emission is the superposition of
both thermal (modelled by a Raymond-Smith plasma) and possibly
non-thermal (modelled by a power law) components. 
We discuss the origin of the thermal component in terms of a 
population of SNRs residing in the disk. The origin of the power law component remains uncertain. 
By comparing the spectrum of the diffuse emission in hard
X-rays ({\sl RXTE}) and soft $\gamma$-rays ({\sl OSSE}),
we find indications that the emission in the two bands are related.
We discuss its origin in terms of both discrete hard
X-ray sources and radiation mechanisms such as  
non-thermal bremsstrahlung from cosmic ray
electrons and protons, and inverse Compton scattering.

The plan of this paper is as follows. 
In \S 2, we describe the
observations.  In \S 3, we present the results of our spatial and spectral 
analysis of the data. \S 4 is devoted to the
discussion of the results, and their implications for the origin of the
emission. Finally in \S 5, we present our conclusions.

\section{OBSERVATIONS} 

The Galactic ridge observations were performed in September of 1996, with the PCA 
instrument onboard the {\sl RXTE} satellite. The PCA (Jahoda et al. 1996) 
has a total collecting area of 6500~${\rm cm^2}$, an energy range of
$2-60$~keV, and energy resolution of $\sim 18\%$ at 6~keV. The collimator
field of view is approximately circular ($2^\circ$ diameter) with  
FWHM of $1^\circ$. 

The observations consist of 14 long scans (from $ 56^\circ$ to $-44^\circ$ 
in Galactic longitude) 
parallel to the Galactic plane up to $\pm 1^\circ\! .5$ and separated 
by $0^\circ\! .25$ in Galactic latitude. The Galactic mid-plane (i.e. 
$b=0^\circ$) was scanned twice. Fourteen shorter scans 
at higher latitudes parallel to the plane ($-15^\circ < l< 15^\circ $; 
$-4^\circ < b < -1^\circ\!.5$ and
$1^\circ\!.5 < b < 4^\circ$) were also performed. 
The speed of scans  ranged from $\sim 0^\circ\!.6$ per minute for the 
shorter scans to $\sim 1^\circ\!.7$ per minute for the longer ones.

We constructed maps of the Galactic ridge X-ray emission from 
these scans in 3 energy bands.
The results are shown in Figure~1.
After subtraction of the instrument background (described in \S~3.2), the net counts were binned
into $0^\circ\! .25 \times 0^\circ\! .25$ pixels for the long scans ($b\le 
| 1^\circ\! .5|$) and $0^\circ\! .25 \times 0^\circ\! .36$ pixels for the
shorter ones. 
The bins were then normalized according to exposure time. The net results are 
reported as count rates (${\rm cts \, s^{-1}}$) in each pixel.
To smooth the images, we have interpolated them using IDL (Interactive
Data Language). Since
the pixel size for the long and short scans were different, each 
image was broken into 3 pieces: one consisting of long scans, and two
others consisting of short scans at larger positive and negative
latitudes. Each piece was then interpolated separately. 
Some of the scans at larger $|b|$ were occulted by the Earth.
Therefore, we have only shown regions that were devoid of data gaps and
could be smoothly interpolated.  
In order to emphasize the Galactic ridge diffuse emission in the images,
we have used a color table that saturates the emission from discrete
sources. Furthermore, we did not attempt to de-convolve the detector's
response function with the light curves. As a result, discrete sources
appear as brown extended regions. 
For the first time, it is clear from these maps that the longitude extent of the diffuse emission in the hard X-ray band 
(above 10~keV) is similar to that in the softer 2-10~keV band. 
 
\section{ANALYSIS AND RESULTS}  

\subsection{Subtraction of the Contribution of Point Sources 
from the Diffuse Emission Spectrum}

To measure the spectrum of the diffuse X-ray emission from the
Galactic ridge, we must first identify and exclude the contribution
of compact sources within the paths of the scans. This
task was done by an automatic routine which identifies sources with a minimum
detection of $3\sigma$. 
The location of each source was calculated
and compared with a catalogue of known X-ray sources. The time
interval in which each source was within the field of view of PCA 
was then excluded. Finally, the remaining good time intervals 
were inspected manually for detection
of possible new sources. A detection of at least 10~${\rm cts \, s^{-1}}$
($\approx 1$~mcrab) 
above background was interpreted as a discrete X-ray source and the 
time interval corresponding to the excess counts above background
was excluded.  
From hereafter, these sources are excluded in the analysis 
reported in this paper. 
Table~1 lists all the identified sources and their positions which lied
in the path of the scans. 
The faintest detected source in the scans has an intensity of $\sim 1$~mcrab.

\subsection{Background Subtraction}

We used the PCA background estimator program {\it pcabackest} (version~1.4h)  
provided by the {\sl RXTE} GOF (Guest Observer Facility) 
to estimate the background. The total background consists of the
instrument background and the cosmic X-ray background (CXB). 
Variations in the instrument background are modelled as two components. One
component is related to changes in the anti-coincidence rate, while the other is
related to the activation by the SAA (Southern Atlantic Anomaly) passages. 
Our Galactic ridge scans were performed during non-SAA orbits and
therefore the activation induced background was estimated to be
zero by {\it pcabackest}. 
For modeling the latitude distribution of
the diffuse emission (\S 3.3), we subtracted the instrument background
only. For spectral analysis of the diffuse emission (\S 3.4), we have
subtracted both the instrument and the cosmic X-ray background. Since 
Galactic absorption distorts the CXB spectrum at lower energies, this
will result in an underestimate of the Galactic ridge flux at 
energies below $\sim 5$~keV. We describe how we correct the spectrum for this
effect in \S 3.4.2.

\subsection{Latitude Distribution of the Diffuse Emission}

Before performing a spectral analysis of the diffuse emission, 
it is crucial to have some understanding of its spatial 
distribution in the Galactic plane. Determining the
scale height of the emission is particularly important for at
least two reasons. Firstly, it provides important clues for the
origin of the emission.  
For example, the scale height of the emission from inverse Compton scattering
is expected to be larger than that from bremsstrahlung. 
This is due to the
homogeneous nature of the cosmic microwave background radiation and
the large scale height of the optical and infrared photons and cosmic
ray electrons.
Secondly, it reveals the multi-component nature of the
emission. For example, it can be determined whether the emission is
confined to a narrow disk centered on the Galactic mid-plane or whether a
broad halo around the disk also contributes to the diffuse emission. 
 
In order to quantify the latitude distribution of the diffuse
emission shown in Figure~1, we convolve the detector's response function 
with a given spatial model and perform a least squared fit to the data. 
In Figure~2a, we have plotted the averaged 
diffuse emission count rate in the 5-8~keV and
8-35~keV energy bands vs. Galactic latitude.
The count rate is obtained by averaging the counts at a
fixed latitude over the central $30^\circ$ of the Galactic plane    
($-15^\circ <l < 15^\circ$) and then normalizing for exposure time. 
Instrumental background is subtracted.   
We chose the 5-8~keV energy band since above 5~keV the effects of Galactic
absorption is small, and also this band implicitly maps the 6.7~keV iron line. 
The 8-35~keV band was chosen to map the distribution of the hard X-rays. 
We found that the best fit is given by a
2-component model consisting of thin and broad distributions for both
energy bands.
We modelled the thin component with a uniform thin disk, and the broad
component with a disk of Gaussian distribution as a function of Galactic
latitude. 
The contribution from
the thin component is proportional to the distance that our line of sight
traverses through the thin disk. Its value drops rapidly as $|b|$
increases, and reaches a minimum at $|b|=90^\circ$.
The contribution from the Gaussian distribution is integrated along the
line of sight to the edge of the galaxy. 
Moreover, we added a constant component to the model to account for the 
cosmic X-ray background. 
To determine the parameters of the thin and broad components, we constrained the
count rate at large Galactic latitudes to that of the    
cosmic X-ray background estimated from {\it pcabackest} 
and searched the parameter space for the best fit.  
For the $5-8$~keV energy band, the best fit
parameters are found to be $0^\circ\!.1_{-0.06}^{+0.4}$ for the
full width of the thin component, and
$4^\circ\!.1\pm0.9$ for the FWHM of the broad
component at the far edge of the Galaxy ($\chi^2=524$, $\nu=21$). 
The best fit is found when the thin
component is centered at mid-Galactic plane, while the broad
component is centered at $b \approx -0^\circ\! .12 \pm0.15$.
In Figure~2b, we have deconvolved the detector's response function and
plotted the actual contribution of each component from the model in the
$5-8$~keV band. 
For the $8-35$~keV band, the best fit parameters are
$0^\circ\! .1^{+0.3}_{-0.09}$ and $4^\circ\! .8^{+2.4}_{-1.0}$
for the full width of the thin component and FWHM of the broad component
at the far edge of the Galaxy, respectively ($\chi^2=209$,
$\nu=21$). The best fit is obtained when both components are centered
at $b \approx -0^\circ\! .03 \pm 0.15$.
Error bars indicate $90\%$ confidence limits.
(Since the fits are poor, the confidence limits are derived by setting
$\Delta \chi^2= 2.7\chi^2/\nu$.)
We notice that a survey of the latitude distribution of the emission
in the $50-600$~keV with {\sl OSSE} yields a broad distribution that is
approximated with a Gaussian distribution
of $5^\circ\! .4\pm 0^\circ\! .5$~FWHM with no significant evidence of
energy dependence (Purcell et al. 1996). However, in our survey,
the detection of the thin disk component is statistically
significant. If we fit
the  emission in the $8-35$~keV band with a single Gaussian component,
the $\chi^2$ of the fit will increase from $201$ (for the
2-component model) to $467$. 
It is likely that the detection of the thin
component may not be possible with {\sl OSSE} due to its large field
of view.  However, we cannot determine the functional
form of the thin component other than deriving limits on
its width, due to the fact that the upper limit is smaller than the field
of view of the PCA. As for the broad component, we point out that it
is most likely related to the one reported by Iwan et al.
(1982) from the measurements taken with the {\it HEAO} A-2 instrument.
The scale height of the diffuse emission from
that study was determined to be greater than $1.5$~kpc.

From the reduced
$\chi^2$ reported above, it is clear that the fits are poor.
This is partly the result of the asymmetry of the overall
profile of the diffuse emission with respect to the Galactic mid-plane.
From Figure~2a, it also appears that at some latitudes the count rates
substantially deviate from that of the model. We have inspected the
source excluded data,
but they do not appear to be contaminated with any
sources above our detection level. Hence, we conclude that
additional spatial components at certain latitudes
corresponding to localized features of the diffuse emission
are needed to produce a better fit.
However, the derived parameters provide a good global description of
the spatial distribution of the diffuse emission in the Galactic plane.

\subsection{Diffuse Emission Spectrum and its Variation as a Function of
Galactic Latitude }

To determine the spectral 
variation of the diffuse emission from the Galactic ridge 
as a function of Galactic latitude, 
we performed separate spectral analysis on the data from 
the central ridge (R1), the northern ridge (R2), and the southern ridge (R3).
The coordinates of each region are given in Table~2.
We will discuss the results of spectral fitting for the central ridge 
first. Spectral analysis for regions R2 and R3 will be  
presented in \S 3.4.3. 

\subsubsection{Central Ridge Spectrum in the $10-35$~keV band} 

Since Galactic disk absorption distorts the spectrum of both the 
Galactic ridge emission and the cosmic X-ray background (CXB) at lower
energies, we analyse the data above 10~keV first, where
absorption is negligible for the expected Galactic
column densities. This strategy
will allow us to constrain the component(s) that dominate the spectrum in the
hard X-ray band. We will then use this derived information to model the
spectrum at lower energies. 

The presence of a hard power law tail above 10~keV has been reported
from combined observations with {\sl Ginga} and {\sl Welcome-1} 
(Yamasaki et al. 1997) and {\sl OSSE} (Skibo
et al. 1997). Here, we search for this component by fitting the {\sl RXTE} data
in the $10-35$~keV energy range with a single power law component. 
The best fit yields a photon index of $\Gamma=2.1^{+0.2}_{-0.3}$
with a normalization of $k=8.5^{+8.5}_{-4.0} \times 10^{-3}\,{\rm photons
\,keV^{-1} \,cm^{-2}}$ at 1~keV
($\chi^2/\nu=56.5/44=1.3$). The surface brightness for the $10-35$~keV
band is $4.9 \times 10^{-8} \, {\rm erg \, s^{-1} \, cm^{-2} \,
sr^{-1}}$. Figure~3 shows the unfolded spectrum. 
A thermal bremsstrahlung for the data above 10~keV  
requires a temperature of $20^{+11}_{-5}$~keV. However, we prefer the
power law model since it is not clear how the hot plasma 
can be bound to the Galactic disk (the gravitational potential of the
disk is $\sim0.5$~keV). Furthermore, the presence of the power law at soft
$\gamma$-ray energies has been established with {\sl OSSE}.  
It is likely that the emission in the hard X-ray and soft $\gamma$-ray
both modelled by a power law are related. We will discuss this further in
\S 3.6.

\subsubsection{Central Ridge Spectrum in the $3-35$~keV band}

Let us now extend the data analysis to lower energies.  
Below 10~keV, the presence of a $6.7$~keV emission line from
He-like iron was discovered with {\sl Tenma} (Koyama et al. 1986).
Further extensive study
of the ridge spectrum with the high resolution capability of {\sl ASCA}
has revealed the presence of lower energy emission lines from Mg, Si, and
S (Kaneda et al. 1997).  These findings have
motivated the idea that part of the emission has thermal origin.
Furthermore, we use the results of the previous section where the
presence of a power law tail
in the spectrum above 10~keV has been established.
Hence, we model the data with both Raymond-Smith (RS) 
plasma and power law components. Notice from Figure~2 that the 
emission in the central ridge (R1) arises from both the thin disk and 
the broad components.  
However, since the spectral decomposition of the broad and thin 
components is complicated and not well constrained with current 
statistics,  we model the averaged effect of the two components. 

Before modeling the data, we should first consider the 
effect of Galactic disk absorption on the final spectrum. 
Since we have subtracted the high latitude cosmic X-ray background from
our data, the Galactic ridge flux at lower energies may have been
underestimated. 
This is because for expected Galactic column densities, 
absorption reduces the CXB flux 
mostly below $\sim 10$~keV, but will leave the spectrum above 
10~keV unchanged. Therefore, in order to compensate for this effect,
we subtract the high latitude CXB model from
our spectral model but add the absorbed CXB model 
(had it been absorbed by the Galactic disk). 
The spectrum of the CXB below 15~keV was modelled from
{\sl RXTE} high latitude observations. The spectrum was
found to have a power law
shape of photon index $\Gamma=1.47$ with normalization
$k=3.6 \times 10^{-3}
\,{\rm photons \,keV^{-1} \,cm^{-2}}$ at 1~keV corresponding to a
surface brightness of $5.7 \times 10^{-8} \, {\rm erg \, s^{-1} \,
cm^{-2} \, sr^{-1}}$ in the $3-10$~keV.  
Notice that since {\sl RXTE} is not designed to separate 
instrument background from the
cosmic X-ray background, some residual counts inherent to the detectors'
background may exist in the CXB model at higher energies, 
but its effect on our modeling is negligible. 
In effect, we add 
the term $\bigl(-kE^{-\Gamma} (1-\exp^{-N_H\sigma(E)})\bigr)$ to our model 
where $kE^{-\Gamma}$ is the CXB model and 
$N_H$ and $\sigma(E)$ are the Galactic column density and the
photoelectric cross section as a function of energy, respectively. 
However, the precise value of the Galactic hydrogen
column density as a function of position and its averaged effect over
$90^\circ$ in Galactic longitude (the span of the central ridge
studied here) are not well known.
Therefore, while we freeze the parameters of the
CXB model in our analysis, we fit for the
Galactic column density along with the rest of the model
parameters. This correction is applied to all models we present hereafter.  

The best fit parameters for the model described above are presented in Table~3.
The data and unfolded spectrum are shown in Figure~4. 
From these results, on average, the emission can be explained
by a Raymond-Smith plasma of temperature $\sim 3$~keV and a
power law component with photon index $\Gamma$ 
$\sim 1.8$ which dominates above
10~keV. In the $3-10$~keV band, 54\% of the flux is from the thermal 
Raymond-Smith plasma component. Above 10~keV, the power law component
dominates by providing 95\% of the flux. 
Confidence contours for $kT-\Gamma$ are shown in Figure~5.  
The average column density derived from this model is
well within the expected Galactic values. It may be argued that
the column density $N_H$ absorbing the 
CXB component may be larger than that absorbing the RS plasma and power law
components since in the case of CXB the entire  
material between us and the edge of
the Galaxy acts as the absorber (i.e. longer path length). 
Had we frozen the value of the column
density absorbing the CXB component at an upper limit of $3 \times 10^{22}\,{\rm
cm^{-2}}$, the key fit parameters would change to
$N_H=1.3^{+0.8}_{-0.6}\,{\rm cm^{-2}}$, $\Gamma=1.85\pm0.15$, and 
$kT=2.9^{+0.4}_{-0.3}$~keV ($\chi^2/\nu=112.9/60=1.9$). These
are all within the error bars of the values presented in Table~3.    

We point out that these results do not indicate that the spectrum
measured with {\sl RXTE} 
is inconsistent with that from {\sl ASCA}. We modeled the {\sl RXTE} data from 
region R1 in the $3-10$~keV with the same components with which
{\sl ASCA} data from the Scutum arm region was modeled (Kaneda et al.
1997), namely 2 thermal components of temperature $\sim 0.8$ and $\sim 7$~keV,
respectively. The fit was acceptable with $\chi^2/\nu=13.5/10=1.35$.  
However, when we extend this model to higher energies, the fit
is poor ($\chi^2/\nu=141.9/58=2.4$).  
This is mainly due to the observed count rate being 
greater than that of the model 
at higher energies, and can be easily seen from Figure~6.  
In the case of {\sl RXTE} analysis, the additional data at higher energies 
allows us to constrain the component that dominates in the hard X-ray band.
This in turn, affects the results of spectral modeling at lower energies. 

Finally, we present a slight variation of our model in which the 
power law is flattening below some energy between $10-100$~keV. The
motivation for this model is two-fold. First, unusually large power is
required to be injected into the galaxy in order to explain the
presence of the power law tail via non-thermal 
bremsstrahlung process. We will discuss the
physical reasons for this in \S 4.2. Second, there is evidence for the
flattening of the slope at lower energies (see \S 3.6 where we discuss
simultaneous fitting of the spectrum with that from {\sl OSSE}). 
But for now, we will only present
the model. The attenuation of the power law can be mimicked by
multiplying the power law function by an exponentially absorbing function
$M(E)=\exp(-E_0/E)$.  Table~4 lists the best fit parameters for this
model. Figure~7 shows the model and unfolded spectrum. Approximately
58\% of the flux in the $3-10$~keV is due to the thermal component in
this model.   

\subsubsection{Spatial Variation of the Spectrum as a Function of Galactic
Latitude} 

Let us now examine the spectrum of the diffuse emission from the ridge at
larger Galactic latitudes. 
From Figure~2, it can be seen that the broad spatial 
component dominates the emission at larger latitudes.
We used the same model discussed in the
previous section to fit the data for Galactic regions R2 and R3
(Table~3). The results of the fits are summarized in Tables~5 and 6,
respectively. Since, the broad component is expected to be less 
absorbed than the thin
disk component, it is reasonable that the average values of the hydrogen column
density for these regions have decreased compared to that of the central
ridge.  
The key fit parameters (namely $kT$ of the RS component and the photon index
of the power law)
have remained  within error bars of those derived for the central ridge
indicating similar physical origins at higher latitudes. 
The average flux in the $3-35$~keV band from both regions is 
about 82\% of that from the central ridge after removing the effect
of absorption. 

\subsection{X-ray Luminosity of the Galactic Ridge} 

Let us now estimate the X-ray luminosity of the diffuse emission for 
the Galactic disk in the $2-10$~keV and $10-60$~keV bands.  
In reality, this is a complex problem since it depends on the geometry
and spatial distribution of the
emitting and absorbing regions. A rigorous treatment of this problem is
outside of the scope of this paper. However, we proceed to estimate the
luminosity by making a few simple assumptions.
From our spatial modeling of the latitude distribution of the emission 
(\S 3.3), we found that 2 spatial components (i.e. a thin disk and a
broad component) contribute to the emission. 
Since the flux at $|b|> 1^\circ$ is dominated by the broad component 
(as seen from Figure~2), we will first estimate the average 
volume emissivity of the broad
component from spectral models of regions R2 and R3. 
From the models presented in Tables~5 and 6, 
we find an average flux of $2.4 \times 10^{-11} \, {\rm
erg \, s^{-1} \, cm^{-2}}$ and $1.9 \times 10^{-11} \, {\rm
erg \, s^{-1} \, cm^{-2}}$ in the $2-10$~keV and $10-60$~keV bands,
respectively,  after removing the effect of absorption. 
Dividing the flux by the {\it effective}
solid angle of {\sl RXTE} ($2.97 \times 10^{-4}$~sr)
yields an average surface brightness $B$ of
$8.1 \times 10^{-8} \, {\rm erg \, s^{-1} \, cm^{-2} \, sr^{-1}}$.
For a spherical geometry with uniform emissivity,
the average volume emissivity along the line of sight can be
estimated to be $\bar{\varepsilon}=B ( 4 \pi / l)$
where $l$ is the
depth of the emitting region. This is a good approximation for the
broad component.  We assume that the distance to the
Galactic center and the radius of the Galactic disk are 8 and 10~kpc,
respectively. Choosing an average line of sight of length 16~kpc, 
we find the average volume emissivity  
$\bar{\varepsilon}_{\rm broad}$ to be $2.1 \times 10^{-29} \, {\rm erg \, s^{-1} \,
cm^{-3}}$. The FWHM of the broad component is about
$4^\circ$ corresponding to a disk of height $\sim 1$~kpc. 
This yields $1.8 \times 10^{38} {\rm erg \,s^{-1}}$ for the 
the X-ray luminosity of the broad disk. 
From our spatial model presented in \S 3.3,
the ratio of the luminosity of the broad to the thin disk is about 14. 
The total luminosity of the disk (thin plus broad) in this model is then
estimated to be $\sim 2 \times 10^{38} {\rm erg \,s^{-1}}$ in the $2-10$~keV
band. Similarly, we derive a luminosity of $1.5 \times
10^{38} \,{\rm erg \,s^{-1}}$ in the $10-60$~keV.  

For comparison purposes, we should mention that
Koyama et al. (1986) have reported a luminosity of 
$\sim 10^{38} \, {\rm erg \, s^{-1}}$ from observations with {\sl Tenma}
($2-11$~keV). 
Worrall et al. (1982) have reported a luminosity of $1.4 \times
10^{38} \, {\rm erg \, s^{-1}}$ in the $2-10$~keV band with HEAO-1, while
Warwick et al. (1985) have reported a luminosity of $\sim 1.7 \times
10^{38}\, {\rm erg \, s^{-1}}$ with {\it EXOSAT} in the $2-6$~keV
band (notice that this band does not include the iron line
emission).
From the latest {\sl ASCA} survey, Kaneda et al. (1997) report a luminosity of
$2 \times 10^{38} {\rm erg \, s^{-1}}$ for
the hard component in the $0.5-10$~keV band where the hard
component is modelled by a bremsstrahlung of temperature 7~keV.
The effect of absorption has been taken into account in all the previously
reported estimates.

\subsection{Extension of the Diffuse Emission Spectrum from hard
X-ray to the Soft $\gamma$-ray Band} 

In the previous sections, we established the
presence of a hard power law tail above 10~keV.
Recent diffuse emission measurement of the Galactic plane at longitude
$95^\circ$ with {\sl OSSE} in the $50-500$~keV band also reveals that 
the spectrum can be fitted with a
power law model ($\Gamma=2.6\pm0.3$; Skibo et al. [1997]). 
If the spectrum extends 
smoothly from the hard X-ray to the soft $\gamma$-ray regime, it
means that the same physical processes are dominating the emission 
in a wide bandpass. Unfortunately {\sl RXTE} data at Galactic longitude
of $95^\circ$ are not available yet. 
However, assuming that the spectral shape of the diffuse emission
is the same as a function of galactic longitude 
for $|b|\le2^\circ$ (Purcell et al. 1996), and that only the normalization varies, 
we attempt a simultaneous spectral fit to the spectrum of the diffuse
emission from both {\sl RXTE} and {\sl OSSE}. We use the 
{\sl RXTE} data from the central ridge (R1) above 10~keV 
and simultaneously fit it with  
the data obtained with  {\sl OSSE} at Galactic  
longitude of $95^\circ$. The 
results of the fit appear in Figure~8. The fit yields a power law photon
index of $2.3\pm0.2$ which is within the errors bars of the values derived from
either {\sl RXTE} above 10~keV ($\Gamma=2.1^{+0.2}_{-0.3}$) or {\sl OSSE} 
($\Gamma=2.6\pm0.3$) alone. 
The normalization for the {\sl RXTE} and {\sl OSSE} data are 
$1.5^{+1.2}_{-0.6} \times 10^{-2}$ and $0.19^{+0.32}_{-0.12}$~${\rm photons
\, s^{-1} \, cm^{-2}}$ at 1~keV, respectively. 
The fit is acceptable with a total $\chi^2$ of $147.6$ for
118 degrees of freedom. 
In order to compare the normalizations of the power law component derived
from  {\sl RXTE}  and {\sl OSSE}, we must
calculate the flux per steradian 
(${\rm photons \, s^{-1} \, cm^{-2} \, sr^{-1}}$) for
each instrument. However, for extended emission, the measured flux depends
on the field of view of each instrument and the distribution function of 
the diffuse emission. Hence, an {\it effective} solid angle (i.e. the 
convolution of the distribution of the emission with the detector's response
function) should be calculated for each instrument. 
In the case of {\sl OSSE} (fov of $11^\circ\! .4 \times 3^\circ\! .8$~FWHM), 
convolving the soft $\gamma$-ray Galactic diffuse emission (assuming a Gaussian
distribution with $\sim 5^\circ$ FWHM [Purcell et al. 1996]),
and {\sl OSSE} 's triangular 
response function yields an {\it effective} solid angle of
$\sim 1.3 \times 10^{-2}$~sr. 
Similarly {\sl RXTE} 's solid angle is $\sim 3 \times 10^{-4}$~sr. 
Therefore, the normalization parameter for the power law
component from {\sl OSSE} should be multiplied by $2.3 \times 10^{-2}$
to be compared with the normalization derived from  {\sl RXTE}. This yields
a normalization of $4.4 \times 10^{-3} \, {\rm photons \, keV^{-1}
\, cm^{-2} \, s^{-1}}$ at
1~keV for the power law component measured with {\sl OSSE}, which is a
factor of $3.4$ smaller than the normalization of the power law component
from {\sl RXTE} data. This is expected since the  
intensity of the emission seen with {\sl RXTE} decreases away from the Galactic center.

If we assume that the physical origin of 
the power law is the same for the averaged central ridge and at
Galactic longitude $95^\circ$, then the combined {\sl RXTE/OSSE} fit
gives a narrower constraint on the power law slope.
However, the slope at higher energies is noticeably steeper
than that derived for the
spectrum below 35~keV, perhaps hinting that the power law either
flattens or slowly attenuates at lower energies.

\section{DISCUSSION: IMPLICATIONS FOR THE ORIGIN OF THE EMISSION}  

\subsection{Discrete Sources}

We first consider the contribution of unresolved discrete sources to
the diffuse emission spectrum. The important question is whether
there exist classes of unresolved discrete sources with 
appropriate spatial number
densities and similar spectra to that of the diffuse
emission that can comprise the bulk of the spectrum.

In the hard X-ray band (above 10~keV), the number of known Galactic
hard X-ray sources with similar spectra to that of the Galactic ridge diffuse
emission is small (Levine et al. 1984). 
Black hole candidates in their soft state have 
spectra similar to that of the ridge,  
but this is usually seen at high luminosities when
individual sources can be easily detected.  
Neutron star binaries are known to have hard power law spectra with an
exponential cut off around 20~keV. However, only a handful of them are known
to be active at a time in the Galactic disk. 
Also, their luminosities are mostly in the range $10^{36}-10^{38}\, {\rm erg \,
s^{-1}}$ which can easily be detected from the diffuse emission. 
Unresolved pulsars are also thought to contribute to the Galactic ridge
diffuse emission particularly in the $\gamma$-ray band. A recent study,
however, concludes that pulsar contribution to the diffuse Galactic emission
is significant only above 1~GeV (Pohl et al. 1997). 

In the soft $\gamma$-ray band, simultaneous {\sl SIGMA}/{\sl OSSE} observations 
of the Galactic center region has been performed to distinguish the diffuse
emission from that of compact sources (Purcell et al. 1996). 
Comparison of the source excluded 
spectrum from the Galactic center and 
that from regions which are known to have fewer sources 
(i.e. $l=25^\circ$ and $339^\circ$) has revealed that the spectra are similar in
both intensity and spectral index. If the emission is
due to discrete sources, 
10 X-ray sources of flux $\le 5 \times 10^{-3}\,
{\rm photons \, cm^{-2} \, s^{-1} \, MeV^{-1}}$ at 100~keV must be present
in the field of view of {\sl OSSE} to make up the spectrum. 
Because of the lack of knowledge of such class of sources with 
a uniform space density in longitude, it  
is likely that the emission above 10~keV is of diffuse origin. 

Below 10~keV, because of its better spatial resolution, {\sl ASCA}
is far better equipped than {\sl RXTE} to resolve individual 
faint sources. A recent
fluctuation analysis of the surface brightness of the ridge emission
in the Scutum arm region ($l=28^\circ\! .5, \, b=0^\circ$) with {\sl ASCA} 
has concluded that if discrete sources are to make up the entire spectrum, 
their luminosity must be less than $2 \times 10^{33}\,{\rm ergs \, s^{-1}}$
(Yamauchi et al. 1996). This conclusion narrows the field of candidate
class of sources that can make up the diffuse emission spectrum of the
Galactic ridge. 
Among possible class of sources, neutron stars (with X-ray luminosities mostly 
in the range 
$10^{36}-10^{38}\,{\rm erg \, s^{-1}}$)  are unlikely due to their
large luminosities. A likely class
of sources that has been thought to account for the emission is the RS~CVn type
binaries, since they exhibit the iron K line in their emission spectra.
However, a study by Ottmann \& Schmitt (1992) on the contribution of RS~CVn
systems to the Galactic diffuse emission concludes that while they
contribute about 6\% to the total Galactic background, their contribution 
to the Galactic ridge iron line is too small ($\sim 1\%$). 
Therefore, given the strength of the iron line in the
spectrum of the diffuse emission, it seems that unresolved 
RS~CVn sources do not contribute to the bulk of the emission. 
Cataclysmic Variables (CVs) are another candidate that
may contribute to the Galactic ridge diffuse emission.  
They have similar spectra to that of the ridge and their luminosities fall
in the range $10^{31}-10^{33} \,{\rm erg \, s^{-1}}$. 
However, from the analysis of the {\sl ASCA} GIS pointings of the Galactic ridge, 
Kaneda (1997) has shown that in order for CVs to make up the diffuse
emission, their space densities   
would have to be $\sim 7 \times 10^{-4}\,{\rm pc^{-3}}$,
a factor of ~100 larger than the estimated density of CVs (Patterson
1984). 

For these reasons, it is more likely that the emission is due
to diffuse or extended sources.  
A more detailed review of the previous works on the contribution of 
the discrete sources to the ridge spectrum can be found in Kaneda (1997) 
and Yamasaki (1996), and references therein. 
Finally, we point out that with the upcoming launch of {\it AXAF}, a more sensitive
number-flux distribution of Galactic sources and their exact contribution to
the diffuse Galactic background can be determined. 

\subsection{Diffuse Emission}

Based on our measurements presented in this paper, 
we conclude that the diffuse emission 
from the ridge in the broad X-ray and soft $\gamma$-ray band has both 
thermal and possibly non-thermal origins. 
In what follows, we speculate on the origin of the emission, first focusing
on the thermal component and then speculating on the origin of the power law
tail. 

The thermal component
can be modelled with a Raymond-Smith plasma of temperature $\sim 2-3$~keV 
(Tables~3 and 4). 
However, we point out that the spectral resolution 
of {\sl ASCA} GIS or SIS is far superior to that of {\sl RXTE} PCA. 
Therefore, in order to interpret the spectrum accurately,
we need to carefully examine the spectral features of the diffuse emission
below 10~keV using the {\sl ASCA} data. 
For this reason, we refer to the latest investigation of
the diffuse emission from the Galactic ridge with {\sl ASCA} (Kaneda 1997;
Kaneda et al. 1997).
The spectral examination of the {\sl ASCA} data has 
clearly verified the presence of lower energy emission lines
from Mg, Si, and S.
The spectrum has also been carefully examined for the presence of Fe
K~lines. While the emission line from He-like iron was clearly
detected, the presence of the emission lines from neutral and H-like
iron was not firmly established. This is important since the equivalent
width of the H-like iron line sets an upper limit for the plasma
temperature.  
Kaneda et al. (1997) give an upper limit of $\sim 1/6$ of that of 
the He-like iron for
the normalization of the H-like iron line. This implies  
an upper limit of $4-5$~keV for the plasma temperature.
This is in agreement with our results since the emission line from H-like
Fe is expected to be very weak from a $2-3$~keV plasma, supporting our
finding that the thermal component of the emission arises from a gas of 
cooler temperature. 
Therefore, we argue that unlike previous conclusions (Koyama et al. 1986;
Kaneda et al. 1997), the temperature of the 
X-ray emitting gas need not be so high (5-15~keV) in order to produce the
observed spectrum. This interpretation lessens the difficulties toward 
explaining the origin and confinement of the super hot ($>5$~keV) gas. 

Due to the similarities between the characteristics of supernova
remnants and the thermal nature of the diffuse emission spectrum
presented here, we speculate that a collection of supernova remnants
in the galactic disk, below current X-ray and/or radio detection limits,
is responsible for the thermal origin of the diffuse emission.
The idea was originally introduced by Koyama, Ikeuchi, \& Tomisaka (1986,
hereafter KIT86). 
However, in their calculation, they assumed that the temperature of the
emitting gas is about $\sim 7$~keV. In addition to the fact that
no SNR with such high temperature has ever been detected,  
this assumption also requires a large SN explosion rate 
(one per 10 years) which exceeds the expected Galactic rate of 2-3 per
century (Tammann 1982).  
Below, we revisit their argument in order to derive the SN explosion rate
capable of powering the thermal emission from the ridge using the
adjusted temperature of $2-3$~keV for the emitting plasma.

The first step is to estimate the criteria for which a given SNR is below 
radio detection limit. 
Due to absorption in the Galactic disk, the minimum radio surface brightness
for detecting a SNR is $10^{-20}\,{\rm W \, m^{-2} \, Hz^{-1} \,
sr^{-1}}$. Using the empirical relation between the surface brightness
$\Sigma$, and the diameter of the SNR (Milne 1979), and accounting for
the ambient gas density as a function of scale height of the Galactic
plane (Tomisaka, Habe, \& Ikeuchi 1980), KIT86 reduce the criteria to 
\begin{equation}
\Sigma=\Sigma_0 D^{-3.8} n_a^2 < 10^{-20}\, {\rm W \, m^{-2} \, Hz^{-1}
\,sr^{-1}},
\end{equation}
where $\Sigma_0=2.88\times 10^{-14}\,{\rm W \, m^{-2} \, Hz^{-1}\,sr^{-1}}$,
and 
\begin{equation}
D=2R_s=2(0.31 n_a^{-0.2}E_{51}^{0.2} t^{0.4})\,{\rm pc},
\end{equation} 
where $R_s$  is the radius of the SNR in the adiabatic stage described
by the similarity solution (Sedov 1959; Spitzer 1968), and $t$,
$E_{51}$ and $n_a$ are the age (years), total energy of the SNR (in
units of $10^{51}$~erg), and the density of the ambient medium (${\rm
cm^{-3}}$), respectively. Substitution of equation~(2) in equation~(1) yields  
\begin{equation} 
n_a t_3^{-0.55} < 0.11 E_{51}^{0.28}, 
\end{equation}
where $t_3=t/10^3$. 

Assuming that the thickness of the emitting region at the shock is
$\sim R_s/12$, the  
X-ray luminosity of the blast wave shell can be calculated from 
\begin{equation}
L_X=(\pi/3)\varepsilon R_s^3,
\end{equation} 
where $R_s$ is described by equation~(2) and 
$\varepsilon$ is the total emissivity of the
gas and depends on the electron density $n_e$ and the cooling function
$\Lambda$ via the relation $\varepsilon=n_e^2\Lambda(T)$. 
For a strong shock $n_e \simeq 4 n_a $. 
The cooling function $\Lambda(T)$ is 
estimated to be (Raymond, Cox, \& Smith 1976) 
\begin{equation}
\Lambda(T)= 2.2 \times 10^{-19}T^{-0.6},
\end{equation} 
for a gas of $10^5\,{\rm K} < T < 4 \times 10^7 \,{\rm K}$ 
($0.008 \, {\rm keV} < T < 3.3\, {\rm keV}$). 
In Figure~9, we have plotted the condition for the undetectability of a 
SNR in the radio band (eq.~[1]), and the restriction on the temperature of the
blast wave ($2\,{\rm keV} < kT < 3\,{\rm keV}$) in the $n_a-t$ phase space
where the temperature just behind the shock front can be calculated
from $T = (3 \mu/16 k)(dR_s/dt)^2$ and $\mu$ and $k$ are the mean mass
of the gas particle and the Boltzman constant, respectively.
We have also plotted relevant constant luminosity lines. 
From the region of the allowed parameters in this plot, 
we conclude that the average luminosity and age of the contributing
SNRs is expected to be $\sim 5 \times 10^{35}-10^{36} \, {\rm erg \, s^{-1}}$
and $(4-5)\times 10^3$~years, respectively. 
The X-ray luminosity of the entire disk in the $2-10$~keV band 
from the {\it thermal} component (which accounts for approximately 50\% of the
emission)  is estimated to be   
$\sim 10^{38}\, {\rm erg \, s^{-1}}$ (\S 3.4.2). 
This implies that a SN explosion rate of $\sim 3-5$ per century 
is required to power the thermal emission from the entire disk. 
On the other hand, the luminosity of the thermal component of the thin disk
is about $6 \times 10^{36}\, {\rm erg \, s^{-1}}$.
Then a SN explosion rate of $\sim 1$ per century is sufficient to power the 
thermal emission from the thin disk.  

Whether this scenario is viable depends on the number of SNRs that are
actually detected in the X-ray band in the Galactic ridge. 
If the SNRs have an average luminosity of $ 5 \times 
10^{35} \, {\rm erg \, s^{-1}}$, approximately 200 of them are needed to
power the thermal emission from the ridge. 
So far, 215 SNRs have been detected in the radio band (Green 1996).
Based on the prevalence of SNRs among a sample of unidentified Galactic
radio sources, Helfand et al. (1989) have built a model which predicts 
the existence of up to $\sim 590$ SNRs in the Galactic disk in the radio band.
According to their results, these remnants are 
within detection capability of current instruments but are not yet
identified.
One of the goals of the {\sl ASCA} Galactic Plane Survey Project is to
identify SNRs in the Galactic disk (in the region $|l|< 45^\circ$ and
 $|b|<0^\circ\!.8$) that
have already been catalogued in other wavebands, and to discover new ones.
Because of Galactic absorption, remnants with low surface
brightness will be missed in the survey. However, SNRs with fluxes
greater than 10~Jy at 1GHz will be detected with {\sl ASCA} 
provided no bright sources are in the field of view. 
From the most recent {\sl ASCA} survey covering the region
$l=342^\circ-357^\circ$ and $l=1^\circ-16^\circ$ pointed at
$b=0^\circ$, 13 SNR were detected (Yamauchi et al. 1998 and references
therein), where
25~SNRs in radio and 1~SNR in X-ray (discovered with ROSAT) were
already known to exist. 
Based on this trend, it can be estimated that 
approximately half of the radio SNRs have significant emission in X-ray. 
It may then be expected that out of the 215~SNRs catalogued in radio,
approximately 100 of them should be detected in X-ray. 
If, we take the
prediction of Helfand et al. (1989) at face value, this number increases to
$\sim 300$. Ultimately, whether the Galactic SNR population can power the
thermal emission from the ridge depends on the final number of detected SNRs and
their average luminosity when the {\sl ASCA} survey is completed.   

We now turn to the discussion of the origin of the power law tail of  
the spectrum dominating the hard X-ray/soft $\gamma$-ray band.  
Recent analysis
of diffuse emission from the Galactic ridge with {\sl ASCA} (Kaneda et
al. 1997)
indicates that the spectrum can be fitted with a two temperature
non-equilibrium ionization plasma of $kT \sim 0.8$ and 7, respectively.
Two factors lead us to believe that the reported high temperature
component of 7~keV with {\sl ASCA} is in fact a different interpretation of 
the power law component reported in this
paper. Firstly, {\sl ASCA}'s sensitivity above 10~keV decreases substantially and
goes to zero around 12~keV. Thus, distinguishing the power law 
from the hard thermal component in the {\sl ASCA} data 
may not be possible. Secondly, 
extension of the two-temperature model above 10~keV
does not produce a good fit to the data at higher energies (as shown in
\S 3.4.2). On the other hand, we have
unambiguously shown here the presence of the hard power law tail in
hard X-ray data. This leads us to conclude
that the previously reported super hot ($> 5$~keV) component of the ISM, is
most likely due to non-thermal emission with a power law spectrum that dominates
above 10~keV and extends to beyond 100~keV. 
Below, we speculate on the origin of the power law tail. 

One scenario is to interpret the hard X-ray emission as the result of the 
superposition of non-thermal tails from SNRs. Previously 
non-thermal X-ray emission
from SN1006 (Koyama et al. 1995) and RX~J1713.7-3946 (Koyama et al. 1997)
has been reported. Non-thermal $\gamma$-ray emission from Cas~A 
(The et al. 1995) and a high energy tail in its X-ray spectrum which
can be possibly attributed to non-thermal processes (Allen et al.
1997) has also been discovered.  
However, it is not clear if this scenario can account for the bulk
of the emission.
The X-ray luminosity of the non-thermal emission 
discovered is about an order of magnitude smaller than the luminosity of
the thermal component.  
In the case of Cas~A for example (G.~E.~Allen, private
communication), the luminosity in the 2-10~keV band is 
$1.6 \times 10^{36}\,{\rm erg \,s^{-1}}$, while the luminosity in the
10-60~keV is about $2\times 10^{35}\,{\rm erg \,s^{-1}}$. From our
estimates, the luminosity of the ridge is approximately 
$ 2 \times 10^{38}\,{\rm erg \,s^{-1}}$ and $1.5\times 10^{38}\,{\rm erg
\,s^{-1}}$ in the 2-10~keV and 10-60~keV bands, respectively. 
The contribution of the SNRs to the thermal component of the spectrum
was previously addressed in this section. For the power law tail
(interpreted as non-thermal
component), this suggests that approximately 750 SNRs with non-thermal
luminosities of $2\times 10^{35}\,{\rm erg \,s^{-1}}$ are needed to
make up the high energy tail of the spectrum.  
Thus far, substantial number of remnants that
exhibit non-thermal emission has not been discovered.
Therefore, the viability of this scenario awaits further observations. 

X-rays of 10-100~keV can also be produced via inverse Compton
scattering of energetic electrons on the ambient microwave background,
optical, and infrared photons. A power law scattered photon spectrum
with photon index $(p+1)/2$ is
expected if the electron spectrum has energy index $p$. 
However, the exact contribution from this
process is model dependent due to the uncertainty in the interstellar
radiation field and the electron spectrum. Direct measurement of the
electron spectrum is reliable only at GeV energies because of the
effects of solar modulation at lower energies.
Irrespective of existing models for the Compton scattering contribution
and their extrapolation to lower energies, 
Skibo, Ramaty, \& Purcell (1996) have argued that this mechanism does not
produce the bulk of the emission below 100~keV.  
Since Compton scattering of microwave background and star light
photons require electron energies of 100~MeV-10~GeV, the electron
spectrum must turn up somewhere in that energy range. However, electrons
at these energies produce the Galactic synchrotron radio emission at
100~MHz frequencies. But since the observed radio spectrum breaks at
$\sim 100$~MHz in the opposite sense to that of the $\gamma$-ray
emission, they argue that an inverse Compton origin is unlikely.  

A likely possibility for the origin of the power law tail is 
non-thermal bremsstrahlung.  However, bremsstrahlung of  
sub-MeV electrons is highly inefficient due to the fact that 
ionization and Coulomb 
collision losses exceed  bremsstrahlung radiation. 
For example, for a 100~keV electron, 
the radiation yield (i.e. the ratio of energy loss from radiation to 
that from collision) is $\sim 2.5 \times 10^{-4}$ (Berger \&
Seltzer 1964). For a 
10~keV electron, the yield drops to $3.8 \times 10^{-5}$. 
From our estimate, the luminosity of the ridge in the $10-60$~keV band
is  $\sim 1.5 \times 10^{38}
\, {\rm erg s^{-1}}$. An order of magnitude estimate implies that in
order to account for the power law tail via electron bremsstrahlung, a
power of $10^{41}-10^{42}\, {\rm erg \, s^{-1}}$ is required. 
Skibo et al. (1996) have calculated that 
a total power input up to $\sim 10^{43}\, {\rm erg \,s^{-1}}$ is 
required if the tail extends down to 10~keV. 
The total injected power into the Galaxy via supernova explosion is only
$\sim 10^{42}\, {\rm erg \, s^{-1}}$ assuming $10^{51}{\rm erg }$ is released in
a SN explosion every 30 years. 
Several attempts have been made to explain the source of this large power. 
Skibo et al. (1996) have suggested that the 
power is derived from the release of the gravitational potential upon the
passage of ISM through galactic spiral arm compressions. Schlickeiser
(1997) explains the implied presence of large population of sub-MeV 
electrons with the existence of interstellar in-situ reacceleration of
cosmic ray particles by the ambient interstellar plasma turbulence.  
Based on the empirical results of modelling the data in the wide
bandpass of 3-500~keV, 
we suggest that the power law tail should gradually attenuate 
below some energy in the 10-100~keV range (Table~4). 
This speculation is motivated by comparing the spectral fits of the
data above and below 10~keV. Above 10~keV, we found that the spectrum
can be modelled with a power law of photon index $2.3\pm 0.2$ up to 
500~keV. On the other hand, when modelling the data in the $3-35$~keV
band, the power law slope is $1.8\pm 0.1$. 
This interpretation lessens the large power requirement 
for the origin of the power law component via bremsstrahlung and
results in the required injection power to be consistent with that
expected from SN explosion in the galaxy.
Unfortunately, the electron spectrum cannot be measured due
to solar modulation at lower energies. However, our interpretation 
of the data in this way makes a prediction for the shape of the 
electron spectrum at lower energies. 
A detailed calculation of
the electron spectrum in this scenario is out of the scope of
this paper. 

In addition to electron bremsstrahlung, 
we suggest the possibility that {\it inverse} (or proton) 
bremsstrahlung may
also contribute to the diffuse emission from the Galactic ridge. 
In the inverse process, the rapidly moving proton radiates instead of the
electron and the center of  momentum of the
proton-electron system is approximately that of the moving proton. 
Boldt \& Serlemitsos (1969) have shown that significant
radiation is associated with the collisions of suprathermal protons
with ambient electrons (free and bound). 
For the non-relativistic case ($\beta \ll 1$), the spectral
intensity from the bremsstrahlung of an electron with kinetic energy $E$
is the same as that from a suprathermal proton of kinetic energy $(M/m)E$,
where $m$ and $M$ are the rest mass of the electron and proton, 
respectively. For example, an electron with kinetic energy of 10~keV
produces the same spectral intensity as that of a proton with kinetic
energy of $20$~MeV. The scenario is particularly attractive in the context
of particle acceleration and reacceleration in supernova shocks via the
diffusive Fermi process. For many years, the perception has persisted that
supernova remnants are a major source of Galactic cosmic rays (e.g. Lagage
\& Cesarsky 1983). Recently much work has been done regarding the 
production of $\gamma$-rays in the SNRs. Modelling of $\gamma$-ray
production via interaction of the energetic particles with the remnants'
ambient environment through neutral pion decay, bremsstrahlung, 
and inverse Compton scattering is reviewed by Baring (1997).  
One expects that if both electrons and protons are
accelerated with equal velocities in the explosion process, both should
contribute to the bremsstrahlung emission. The ratio of the
bremsstrahlung losses to the collisional losses is roughly the same for 
both electrons and protons since it is only dependent on the velocity of
the radiating particles. 
However, a simple calculation shows that not only the accelerated
protons' lifetime is longer than that of electrons', but also they travel
farther than electrons from their birthplace. To illustrate this, let us
consider bremsstrahlung radiation from an electron of 25~keV. Its
spectral intensity is similar to that from a proton of 50~MeV. However,
the range $R$ of the 25~keV electron in hydrogen is $5.6 \times 10^{-4} \,
{\rm g \, cm^{-2}}$ as opposed to $1.0 \, {\rm g \, cm^{-2}}$ for the
50~MeV protons (Berger \& Seltzer 1964; Barkas \& Berger 1964).
This translates to distances of approximately 100~pc and $0.2$~Mpc 
for the electrons and protons, respectively. The radial distance $S$ from 
the birthplace of the energetic particles which the particles will
travel before they lose all their energy depends on the random
walk step length $\lambda$ via the relation
$S/\lambda = (R/\lambda)^{1/2}$. 
While the value of $\lambda$ is not well known, it is expected to be
much larger than the gyro radii of the particles for the expected Galactic
magnetic fields but somewhat smaller than the height of the Galactic disk.
A good estimate is that it is on the order of the coherence length of the
magnetic fields. As a quantitative value for this length, 
we use the study
of Kim, Kronberg, \& Lancecker (1988) where they probed the
structure of magnetic fields in and around the supernova remnant
OA~184~(G166.2+2.5) using background radio source rotation
measures. Based on their study, they find an interstellar magnetic
irregularity scale of order 100~pc. Using this value, we find 
the total radial distance $S$ from the birthplace of the particles
to be approximately 100~pc and $4.5$~kpc for the electrons and
protons, respectively. In any case, it is clear that the radial distance
travelled by the protons is much larger than that by electrons. 
Likewise, the energy loss lifetime of protons ($E/\dot{E}$) 
is much larger than that
of the electrons (by a factor of $\sim 2000$) since it is proportional to
the mass of the particles.  
From these estimates, one
expects that the proton bremsstrahlung should play a significant role if 
bremsstrahlung is responsible for the presence of the hard power law tail 
in the spectrum. 
This is due to the longer range and energy loss lifetime of the
protons compared to electrons. 
If indeed the birthplace of the energetic electrons and protons are the
sites of supernova explosions in the Galaxy, electron bremsstrahlung
will produce localized hard X-ray emission (i.e. hot spots). On the other
hand, proton bremsstrahlung will cause the radiation to be smoothly
distributed over the Galaxy. This interpretation implies that
SNRs play a dominant role in the energetics of ISM via 
both thermal and non-thermal emission processes. 

Finally, we calculate the hydrogen ionization rate for the ISM 
in this model. 
If non-thermal bremsstrahlung dominates above 10~keV, then from our
model, the average volume emissivity in the $10-60$~keV for the 
ridge is 
$1.9 \times 10^{-29} \, {\rm erg \, s^{-1} \, cm^{-3}}$. 
The average
radiation yield in the $10-60$~keV band is $\sim  10^{-4}$. This
yields a collisional loss rate of $1.9 \times 10^{-25} \, {\rm erg \,
s^{-1} \, cm^{-3}}$. For an average energy loss of 36~ev
($5.8 \times 10^{-11}$~erg) per hydrogen atom ionized in the ISM
(Dalgarno \& Griffing 1958), we find  
an average ionization rate of  
$3.3 \times 10^{-15}n_H^{-1}\,{\rm (atom \, s)}^{-1}$ for the entire
ridge 
where $n_H$ is the average hydrogen density. Previous values derived
from observations have been within 
$10^{-15}-10^{-14}\, {\rm (atom \, s)}^{-1}$
(Dalgarno \& McCray 1972; Reynolds et al. 1973 from $H\beta$
observations). Our estimate is consistent with 
these values for an average $n_H \sim 1\,{\rm atom \, cm^{-3}}$ in the
ISM. 

Independent of $n_H$, from the intensity of Galactic $H\alpha$
background at high Galactic latitudes, Reynolds (1984) has derived 
$(2-7)\times 10^6 \,{\rm s^{-1}}$
hydrogen ionizations (with a ``best fit'' value of  $4\times 10^6
\,{\rm s^{-1}}$) per ${\rm cm^{2}}$ column perpendicular to the 
Galactic disk for the region within $2-3$~kpc of the Sun. 
Our results imply $3.3 \times 10^{-15}\,{\rm s^{-1}\, cm^{-3}}$ hydrogen
ionizations averaged for the ridge.
Since in our model, the thickness of the broad disk is $\sim 1$~kpc, 
our results imply hydrogen ionizations of $ \sim 9.9 \times 10^6 \,{\rm s^{-1} 
\, cm^{-2}}$ of the Galactic disk 
which is larger than the $H\alpha$ estimate by more than a factor of 2. One
reason could be that our estimate is derived by averaging over the central
$30^\circ$ of the plane in longitude where the hard X-ray emission is more
intense, while the rate from $H\alpha$ observation is for regions within
$2-3$~kpc of the Sun. In any case, 
this indicates that the collisional losses associated with the
bremsstrahlung emission from energetic particles 
to produce the hard X-ray tail are more than adequate to 
account for the observed hydrogen ionization rate in the ISM.

\section{CONCLUSIONS}

Our results are summarized as follows: (1)~From the {\sl RXTE} survey of the
Galactic plane, we find that the diffuse emission from the Galactic ridge
has two spatial components: a thin disk of width $\lesssim 0^\circ\!.5$, and a
broad component with a functional form that can be approximated by a 
Gaussian distribution of about $4^\circ$~FWHM. 
Assuming an average distance of 16~kpc to the edge of galaxy, this
translates to a height of 
70~pc and 500~pc for the thin and 
broad disk, respectively. 
(2)~A hard power law tail is clearly detected in the spectrum and
dominates above 10~keV, implying that the emission in hard X-ray 
possibly has non-thermal origin. On the other hand, the detection of
the emission line from He-like iron in the spectrum 
(and also lower energy lines from Mg, Si,
S in the spectrum from {\sl ASCA}) motivates the idea that part of 
the emission below 10~keV has thermal origin.
(3)~The spectrum in the $3-35$~keV band can be well modelled with a 
Raymond-Smith plasma component of $2-3$~keV plus a power law component of 
photon index $\sim 1.8$. The spectrum above 10~keV simultaneously
fitted with {\sl OSSE}'s spectrum can be fitted with a power law of
photon index $2.3\pm0.2$. The change in the power law slope at
lower energies hints at the possibility that the power law either flattens
or gradually  
attenuates below some energy between 10-100~keV. 
(5)~The characteristics of the thermal component of the diffuse emission
resembles that of the SNRs. From this interpretation, we calculate that a
SN explosion rate of less than 5 per century is adequate to power the thermal
emission from the ridge. 
(6)~The origin of the emission in the hard X-ray band modelled by a power
law is uncertain. While unresolved discrete sources are expected to
contribute, the bulk of the emission is expected to be of diffuse
origin. 
One possibility is the non-thermal bremsstrahlung of electrons and
protons which may have been accelerated in the SNR sites. 
Based on the empirical modeling of the data in a wide energy band, we
speculate that the power law either flattens or 
gradually attenuates at lower energies. 
This lessens the large power injection required to explain
the hard power law tail via either electron or proton bremsstrahlung.
It will also make it consistent with the power injected to the Galaxy
via supernova explosion. 
(7)~The hydrogen ionization rate implied from our model
is $3.3 \times 10^{-15}n_H^{-1}\,{\rm (atom \, s)}^{-1}$ averaged over the
central ridge. Generally, values of
$10^{-15}-10^{-14}\, {\rm (atom \, s)}^{-1}$ are expected in the ISM. 
This is an indication that the collisional losses associated with the
bremsstrahlung radiation of energetic particles that produce the hard power
law tail are sufficient to explain the ionization rates observed in the ISM. 

Finally, we expect that simultaneous {\sl RXTE}/{\sl OSSE} observations of 
the diffuse emission from the Galactic ridge will allow the exclusion of hard
discrete sources from the spectrum and will tightly
constrain the hard tail slope that dominates in the hard X-ray/soft
$\gamma$-ray band. This will enhance our understanding of the origin of the
power law tail and the energetics of ISM.

\acknowledgements

We are grateful to Elihu A. Boldt for many valuable discussions, in
particular for bringing the process of proton bremsstrahlung to our
attention and his contribution to the relevant arguments presented in this
paper. We also thank him for a careful reading of the manuscript. 
We acknowledge  G. E. Allen, E. Boldt, K. Ebisawa, S. Hunter, K. Jahoda, 
G. M. Madejski, R. M. Mushotzky, and R. Petre for useful discussions. 
We are grateful to J. G. Skibo for
providing the diffuse emission spectrum measured with {\sl OSSE}. 
The help of Pat Tyler and Seth Digel 
in producing Plate~1 is greatly appreciated. 
We thank the referee, K. Koyama, for many insightful comments that led
to the improvement of this paper. 
A.V. acknowledges support from the
National Academy of Sciences and the National Research Council through
a research associateship at NASA's Goddard Space Flight Center.  

%

\newpage

\begin{figure}
\vspace{7in}
\includegraphics{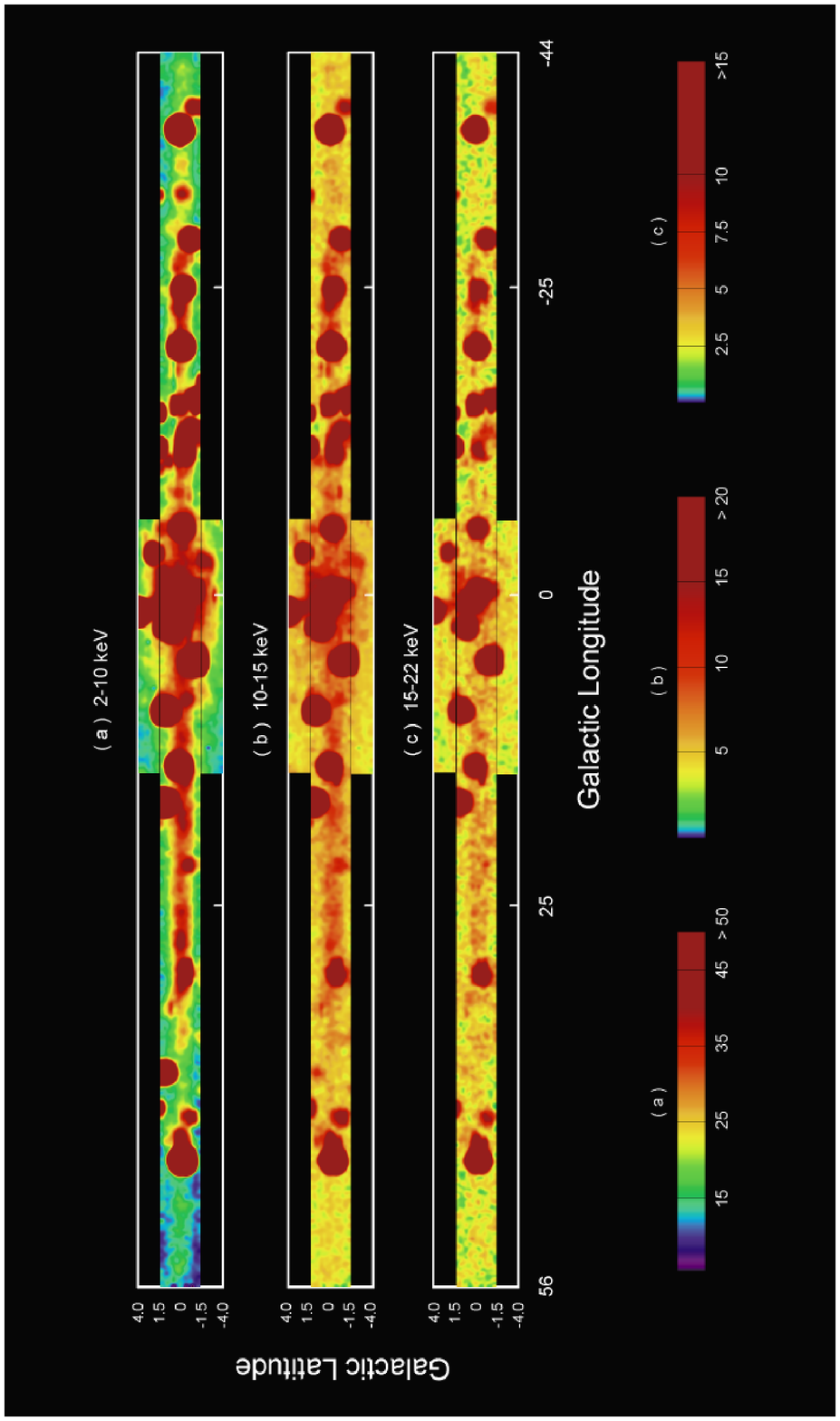}
\caption{
Maps of the Galactic ridge X-ray emission for 3 energy bands of
(a)~$2-10$~keV, (b)~$10-15$~keV, and (c)~$15-22$~keV.
Color bars corresponding to each map appear
at the bottom.  See \S 2 for details.
{\bf Note: A high quality tiff image of this color plate can be 
obtained via anonymous ftp to
ftp://lheaftp.gsfc.nasa.gov/local/LHEA/FTP/pub/valinia/ridge/ridge.tiff. 
For a hard copy, write to Azita Valinia at valinia@milkyway.gsfc.nasa.gov.}
}
\end{figure}

\begin{figure}
\vspace{4in}
\includegraphics{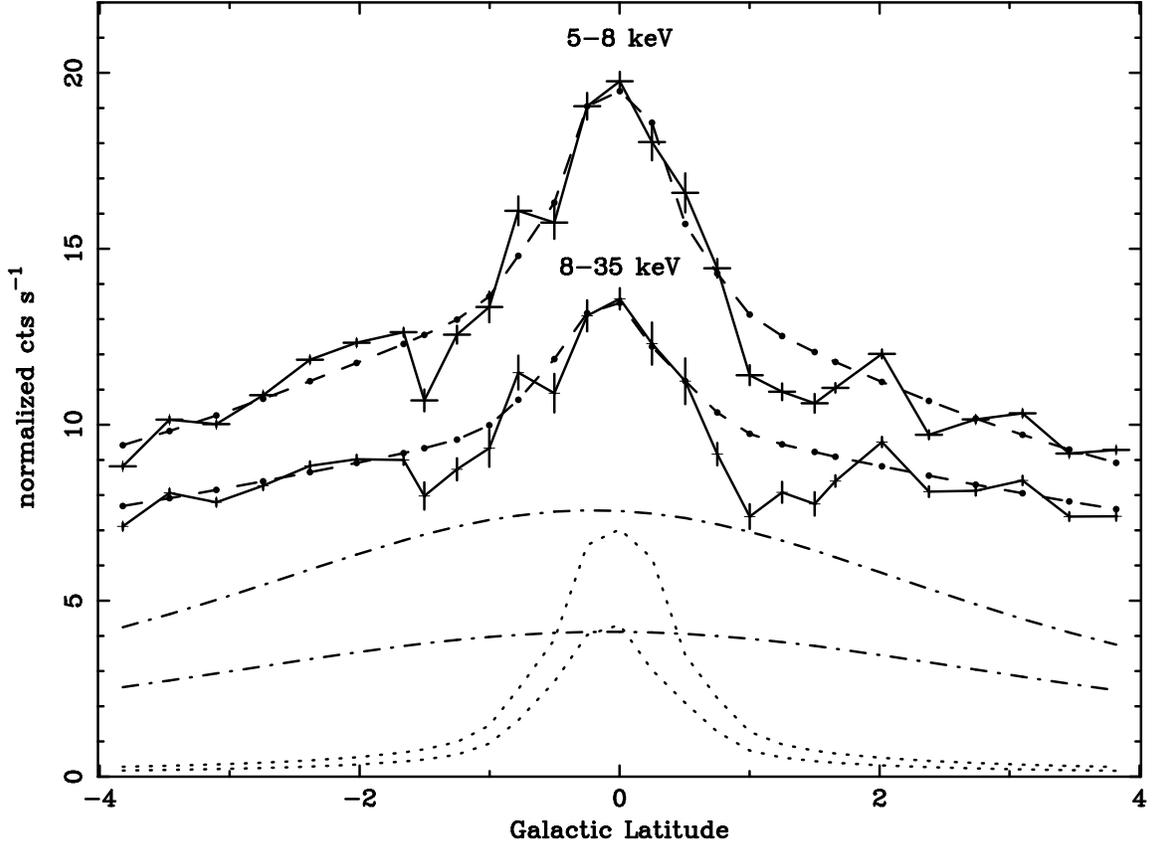}
\caption
{(a) (this page)~Latitude Distribution of the diffuse emission averaged
over
$ -15^\circ < l < 15^\circ $ for the $5-8$~keV and
$8-35$~keV energy bands (solid  curves). Position of the error bars
show the measured averaged count rate
at that latitude. Dashed curves represent the results of the
convolution of a thin plus broad spatial model to the detector's
response function. Dotted and dash-dot
curves show the convolution of the thin disk and broad models,
respectively ($5-8$~keV: upper curves; $8-35$~keV: lower curves).
A constant component (not shown) is included in the model to account for
the
cosmic X-ray background.
See \S 3.3 for parameters of the fit.
(b)~(next page) Contribution of the thin and broad spatial components
to the 5-8~keV flux as a
function of Galactic latitude in arbitrary units. Dotted curve shows the
combined contribution of both components.
}
\end{figure}

\newpage

\begin{figure}
\vspace{4in}
\includegraphics{rfig2b.ps}
\end{figure}
 
\newpage

\begin{figure}
\vspace{4in}
\includegraphics{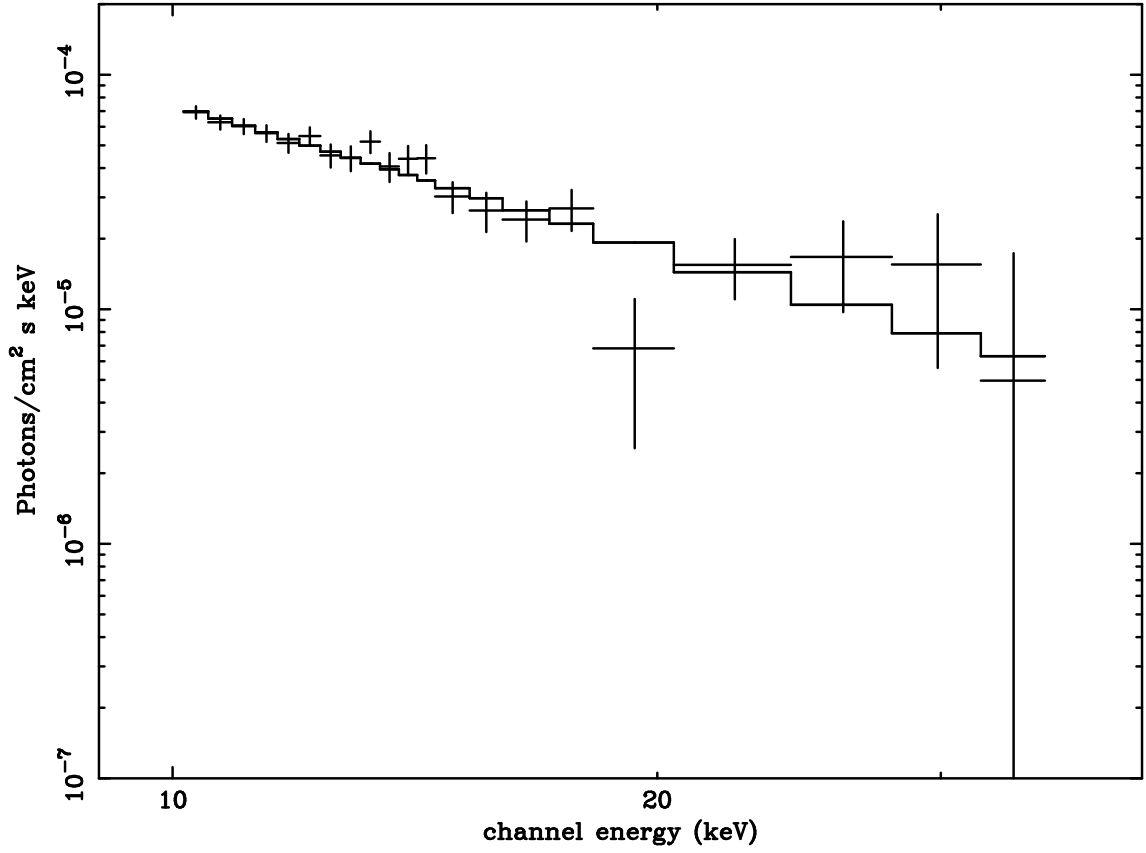}
\caption
{Unfolded spectrum and best fit model (power law)
for the central region of the ridge (R1) in the
$10-35$~keV band.}
\end{figure}

\newpage

\begin{figure}
\vspace{4in}
\includegraphics{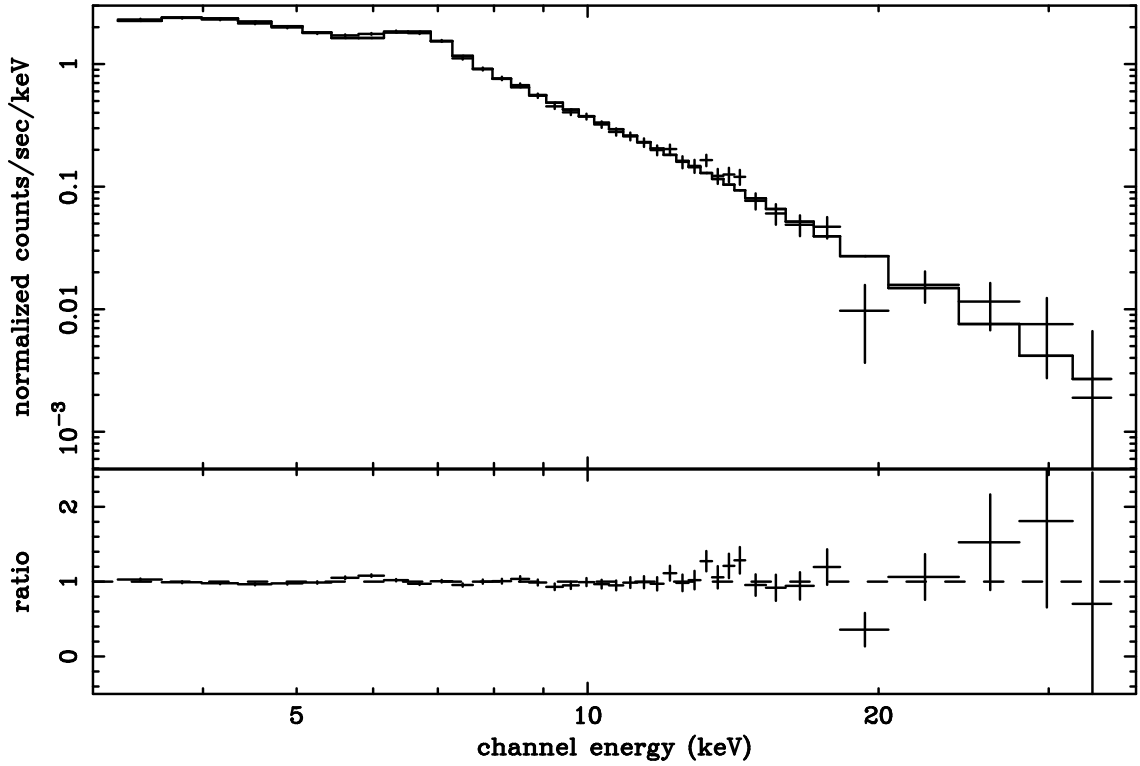}
\caption
{Results of modeling the spectrum from the central region
of the ridge (R1)
in the $3-35$~keV energy band (see Table~3 for model parameters):
(a)~(this page) Data and folded model (b)~(next page)Unfolded spectrum and model.
Contributions to the model
from various components are also shown.
}
\end{figure}

\newpage

\begin{figure}
\vspace{4in}
\includegraphics{rfig4b.ps}
\end{figure}

\newpage

\begin{figure}
\vspace{4in}
\includegraphics{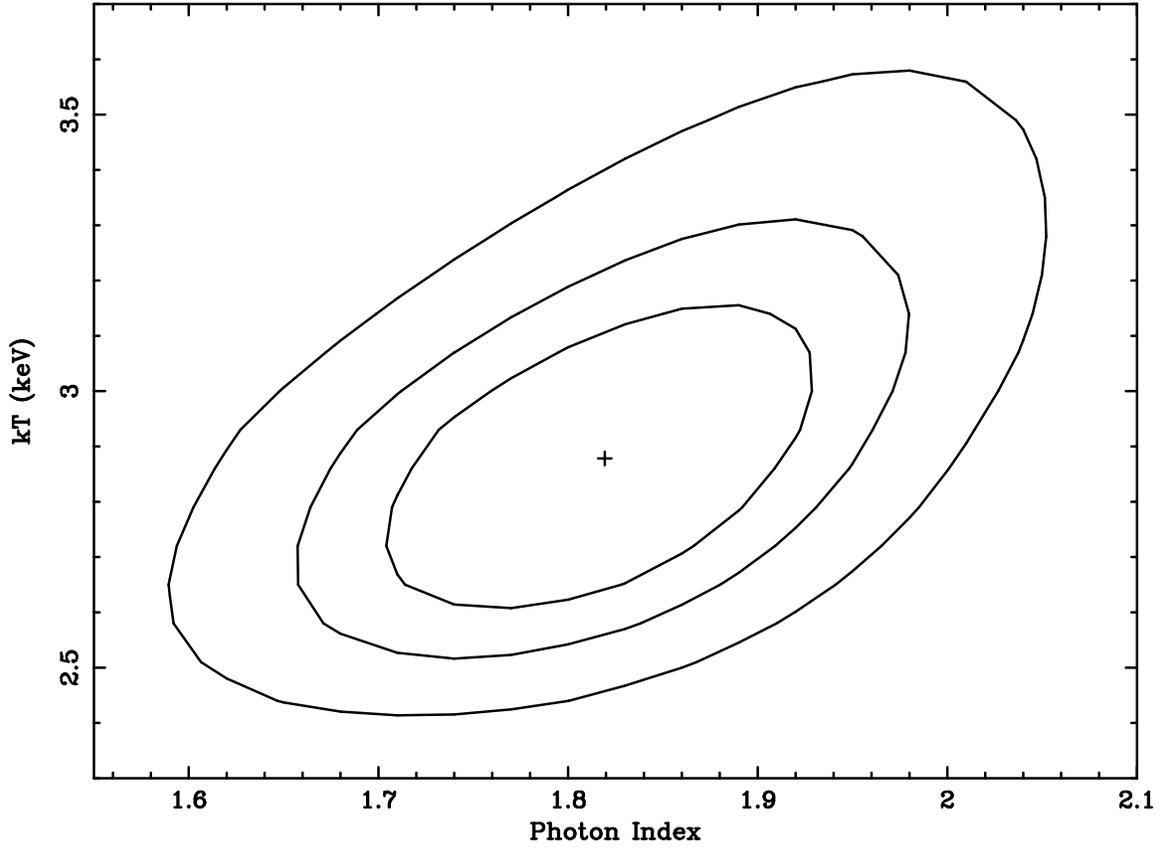}
\caption
{Confidence contour plot of Raymond-Smith plasma temperature vs.
power law photon index for the model presented in Table~3 and Figure~4.}
\end{figure}

\newpage

\begin{figure}
\vspace{4in}
\includegraphics{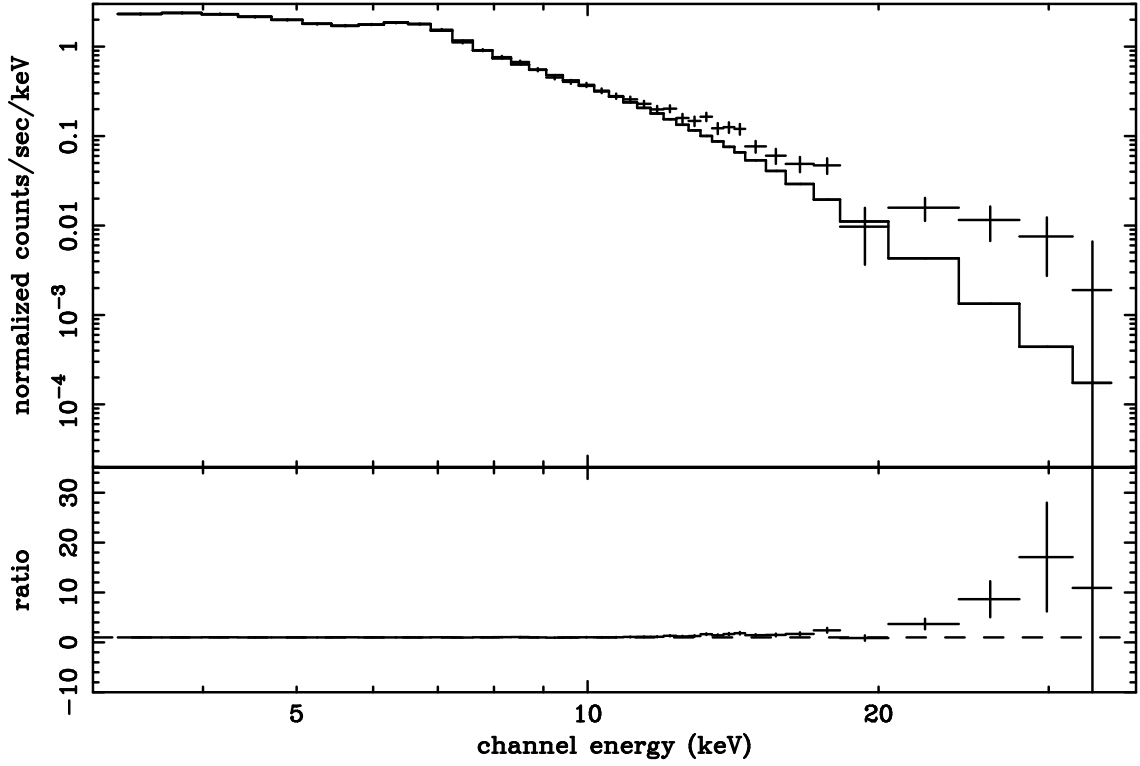}
\caption
{RXTE data in the 3-35~keV band folded with the best fit model
derived from analysis of ASCA data in the $2-10$~keV band (Kaneda et al.
1997).
}
\end{figure}

\newpage

\begin{figure}
\vspace{4in}
\includegraphics{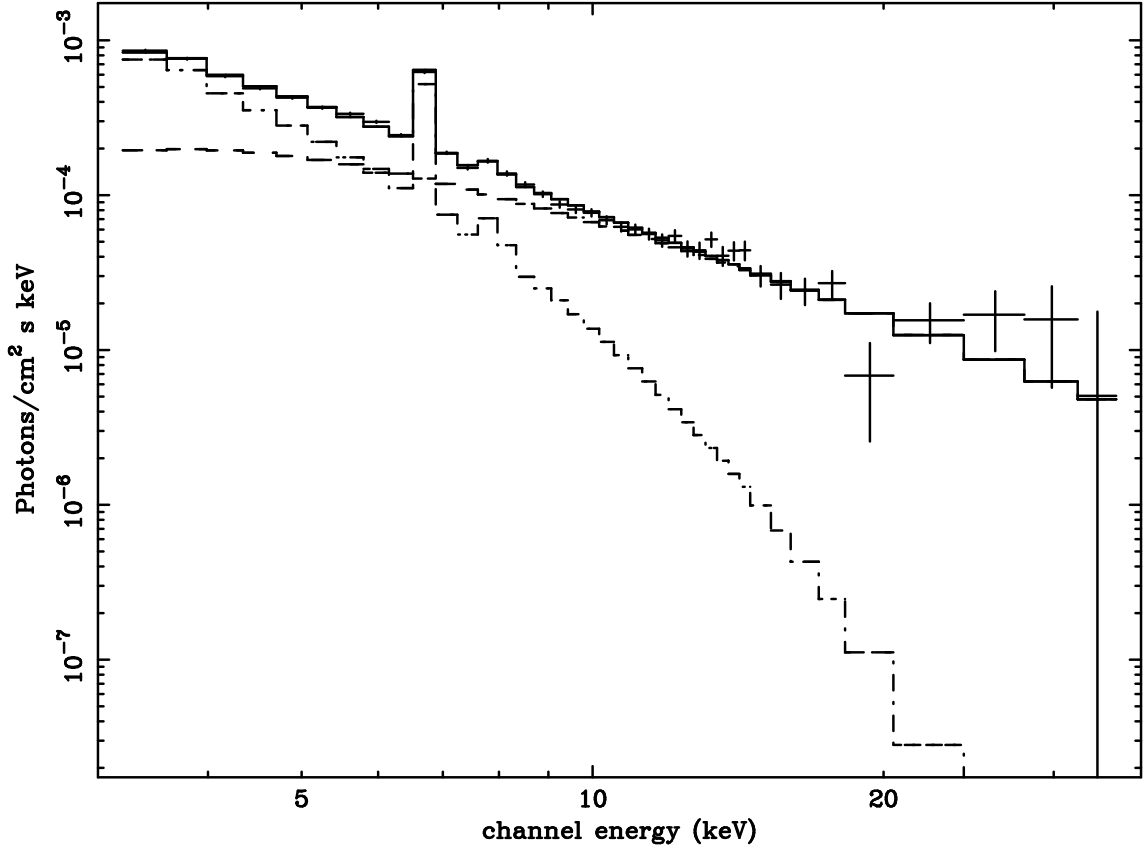}
\caption
{Unfolded spectrum and model (presented in Table~4)
for the central ridge (R1). The power law is multiplied by an
exponentially absorbing function to mimic the flattening of the
power law at lower energies.
}
\end{figure}
 
\newpage

\begin{figure}
\vspace{4in}
\includegraphics{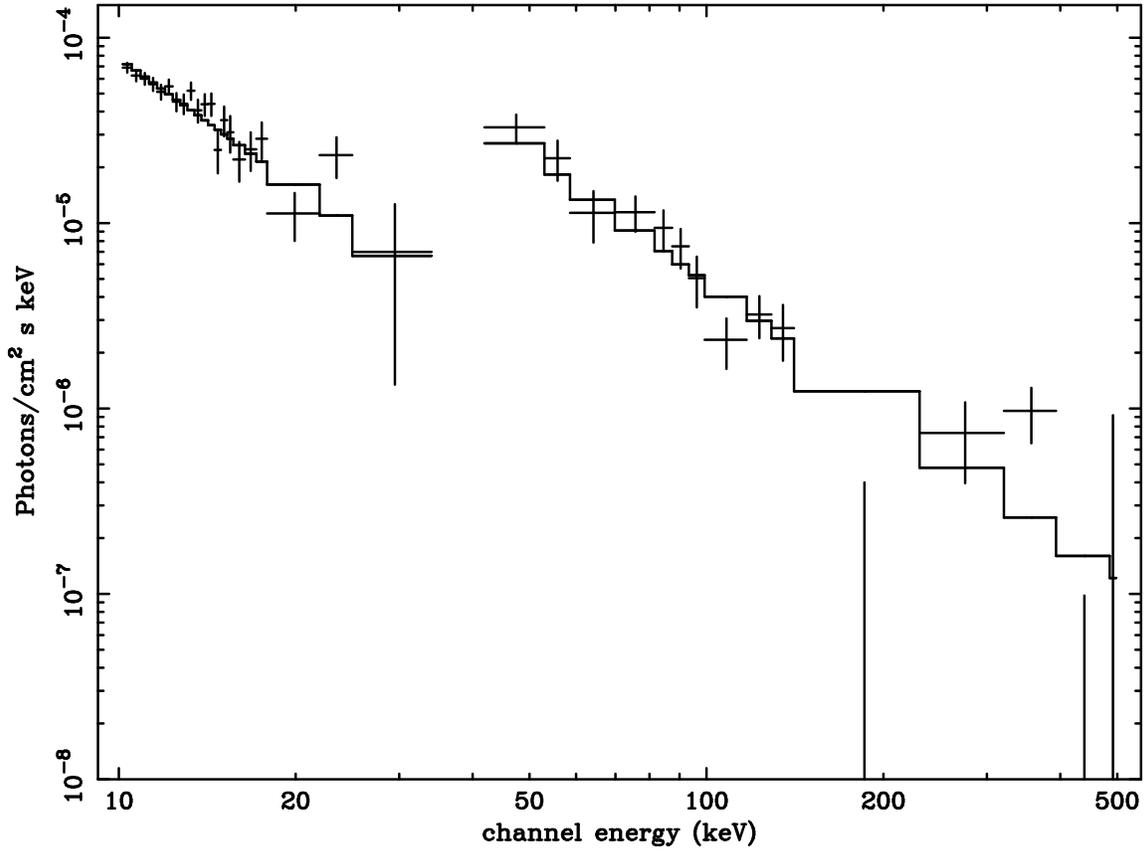}
\caption
{Simultaneous fit to the diffuse emission spectra from $l=95^\circ$
obtained with {\sl OSSE}, and the central region of the ridge (R1)
obtained with {\sl RXTE}. See \S 3.6 for details.
}
\end{figure}

\newpage

\begin{figure}
\vspace{4in}
\includegraphics{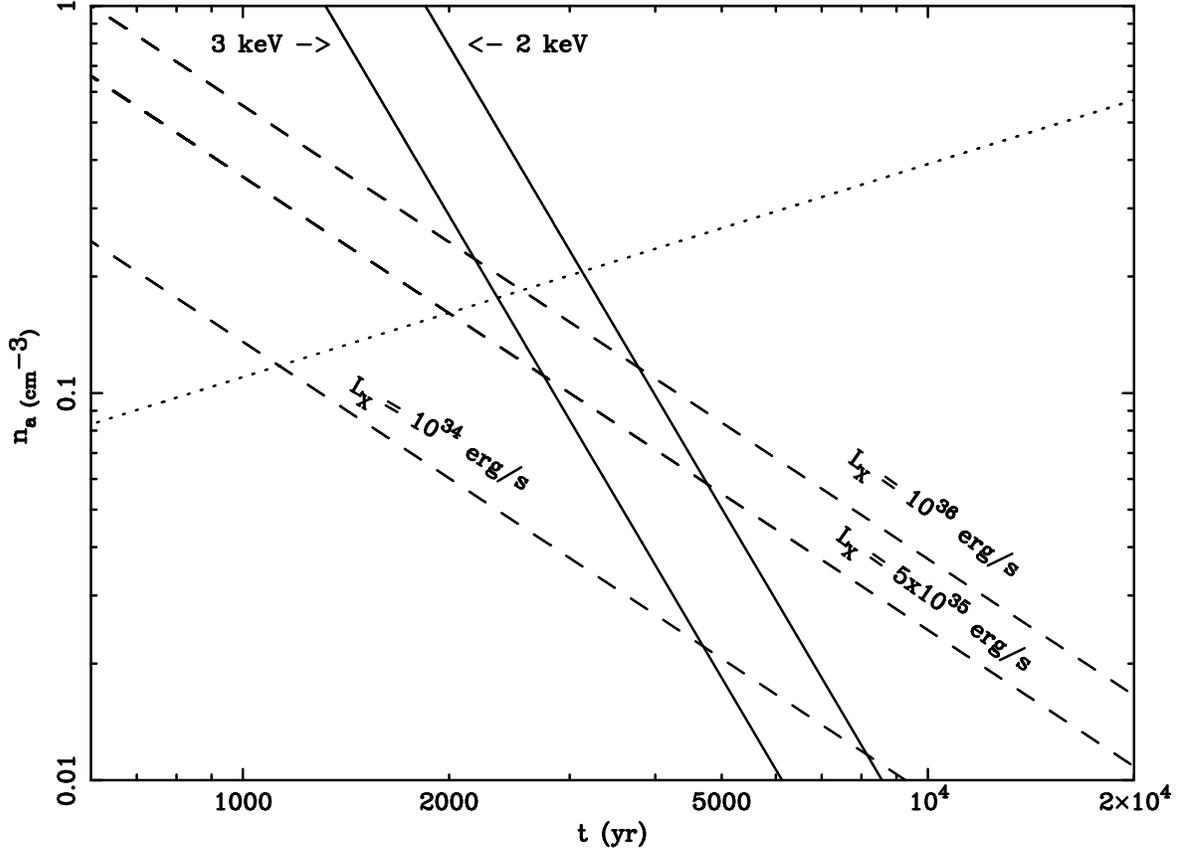}
\caption
{Constraints on the blast wave temperature (solid lines) and SNR
detectability in radio (dotted line) are plotted in the ambient
density-SNR
age phase space. Dotted line gives the condition of the surface
brightness of
the remnant to be $10^{-20}\, {\rm W \, m^{-2} \, Hz^{-1}\,sr^{-1}}$
(eq.~[3]). Constant luminosity lines (dashed lines)
are also plotted. The allowable zone from
the argument given in the text is confined to the space between the
two solid lines and below the dotted line.
}
\end{figure}

\clearpage 


\begin{deluxetable}{lcc}
\tablewidth{0pt}
\tablecaption{Detected Sources in the
RXTE Scans of the Galactic Ridge
}
\tablehead{
\colhead {Source} &
\colhead {$l$} &
\colhead {$b$} }
\startdata

J1744-28   &  0.044   &   0.301  \nl
GRS1739-278   & 0.660 &     1.167  \nl
SL1735-269  &  0.785  &    2.400  \nl
K1731-260  &  1.074 & 3.655  \nl
4U1744-26 &  2.294  &    0.793  \nl
GX5-1  &  5.079 &    -1.018   \nl
4U1745-203  &  7.725 & 3.795   \nl
GX9+1   & 9.072  &     1.154  \nl
GX13+1  & 13.518  &    0.108   \nl
GX17+2 &  16.432  &    1.276   \nl
4U1812-12  & 18.033  &    2.362  \nl
G21.5-0.9  &  21.493  &   -0.887  \nl
GR11(1)  &  22.001 &    -0.002   \nl
EXO1846-031  & 29.954 &    -0.917  \nl
G1843+009  &  33.108 &     1.746  \nl
GPS1858+015  &  35.416  &    -1.648  \nl
XTE J1856+053\tablenotemark{*} & 38.258 & 1.266  \nl
EXO1902+054  &   39.33  &   -0.390  \nl
SS433  &  39.695  &   -2.247  \nl
4U1909+07 & 42.089  &   -0.940  \nl
W49B   &  43.275  &   -0.189  \nl
A1907+09  &  43.743  &     0.477  \nl
GRS1915+105 &  45.366  &  -0.219  \nl
MSH15-52  &  320.321  &   -1.161  \nl
CIRCINUSX-1  &  322.117  &    0.038  \nl
4U1553-542  & 327.945  &   -0.857  \nl
2E1614-505  &  332.421   &  -0.422  \nl
1608-522   &  330.926   &   -0.850  \nl
4U1624-49  &    334.914  &   -0.263  \nl
GX340+0  & 339.587 &    -0.079  \nl
2S1702-429   &  343.887   & -1.314  \nl
OAO1653-40   & 344.369  &     0.326  \nl
 NSCO94   &  344.982   &    2.456  \nl
4U1708-408  & 346.327 & -0.928  \nl
GPS1709-396   &  347.300   &  -0.333  \nl
4U1700-37   & 347.755  &     2.173  \nl
GPS1713-388  &  348.394   &  -0.543  \nl
CTB37A/B   & 348.598   &   0.232  \nl
GX349+2  & 349.105 &     2.750  \nl
EXO1722-363  &  351.474 &     -0.548  \nl
A1744-361  &  354.121 & -4.192  \nl
MXB1728-34  &  354.305  & -0.151  \nl
X1730-333  & 354.843 &    -0.159   \nl
TERZAN2  & 356.320  &    2.298   \nl
SL1746-331  & 356.816  & -2.997  \nl
H1741-322  & 357.125  &   -1.607   \nl
R1747-313  & 358.556   &  -2.168  \nl
2S1742-294 &  359.559 &    -0.390   \nl
A1742-289   & 359.930  &   -0.043  \nl
\tablenotetext{*} {Previously unknown source discovered with RXTE. See
IAUC 6504}
 
\enddata
\end{deluxetable}

\begin{deluxetable}{cccc}
\footnotesize
\tablecaption{Spatial Divisions of the Galactic Ridge} 
\tablewidth{500pt}
\tablehead{
\colhead {Region} &
\colhead {Galactic Longitude} &
\colhead {Galactic Latitude} &
\colhead {Exposure Time (ks) }} 
\startdata

R1  &    $-45^\circ < l < 45^\circ $  &   $-1^\circ\!.5  < b < 1^\circ\!.5 $
& 21.5 \nl                       
R2  &    $-15^\circ   < l < 15^\circ $  &   $1^\circ\!.5  < b < 4^\circ\!.0 $
& 12.5 \nl                      
R3  &   $-15^\circ  < l < 15^\circ $  &   $-4^\circ\!.0  < b < -1^\circ\!.5 $
& 13.1 \nl 
 
\enddata
\end{deluxetable}

\begin{deluxetable}{lll}
\footnotesize
\tablecaption{Best fit model parameters for the $3-35$~keV spectrum from
region R1 }
\tablewidth{470pt}
\tablehead{
\colhead {Model Component} &
\colhead {Parameter} &
\colhead {Value} }
\startdata
 
Absorption\tablenotemark{a} &  $N_H$ ($10^{22}\,{\rm cm^{-2}}$) &  
$1.8^{+0.6}_{-0.4}$  \nl
Raymond-Smith & $kT$ (keV)  &  $2.9\pm0.3$  \nl
 &        Abundance (Solar)\tablenotemark{b}  &  $1.0$ \nl
 &         Normalization\tablenotemark{c} &  $1.8 \times 10^{-5}$  \nl   
Power Law  &  Photon Index &  $1.8\pm 0.1$  \nl
  &         Normalization\tablenotemark{c} & $6.2 \times 10^{-5}$ \nl \hline 
 &  $\chi^2/\nu=115.4/60=1.9$  &  \nl

\enddata

\tablenotetext{a} {Absorption of CXB spectrum by Galactic disk has also been
taken into account. See \S 3.4.2  for details.} 
\tablenotetext{b} {Frozen}
\tablenotetext{c} {Unabsorbed flux (${\rm photon \, s^{-1} \, cm^{-2} \,
keV^{-1}}$) at 10~keV} 

\end{deluxetable}
 
\begin{deluxetable}{lll}
\footnotesize
\tablecaption{Same as Table~3 except that the power law is multiplied
by an exponentially absorbing function $M(E)=\exp(E/E_0)$ 
}
\tablewidth{470pt}
\tablehead{
\colhead {Model Component} &
\colhead {Parameter} &
\colhead {Value} }
\startdata
 
Absorption\tablenotemark{a} &  $N_H$ ($10^{22}\,{\rm cm^{-2}}$) &
$2.0^{+1.6}_{-0.6}$  \nl
Raymond-Smith & $kT$ (keV)  &  $2.3\pm0.4$  \nl
 &        Abundance (Solar)\tablenotemark{b}  &  $1.0$ \nl
 &         Normalization\tablenotemark{c} &  $1.4 \times 10^{-5}$  \nl
Power Law  &  Photon Index &  $2.7^{+1.2}_{-0.7}$  \nl
  &  $E_0$  &  $8.5^{+7.9}_{-6.5}$ \nl  
  &         Normalization\tablenotemark{c} & $6.8 \times 10^{-5}$ \nl
\hline
 &  $\chi^2/\nu=112.2/59=1.9$  &  \nl

\enddata
 
\tablenotetext{a} {Absorption of CXB spectrum by Galactic disk has also
been
taken into account. See \S 3.4.2  for details.}
\tablenotetext{b} {Frozen}
\tablenotetext{c} {Unabsorbed flux (${\rm photon \, s^{-1} \, cm^{-2}
\,
keV^{-1}}$) at 10~keV}
 
\end{deluxetable}

\begin{deluxetable}{lll}
\footnotesize
\tablecaption{Best fit model parameters for the $3-35$~keV spectrum
from region R2 }
\tablewidth{470pt}
\tablehead{
\colhead {Model Component} &
\colhead {Parameter} &
\colhead {Value} }
\startdata

Absorption\tablenotemark{a} &  $N_H$ ($10^{22}\,{\rm cm^{-2}}$) &  $0.2^{+0.7}_{-0.2}$
\nl
Raymond-Smith & $kT$ (keV)  &  $3.4^{+0.4}_{-0.6}$  \nl
 &        Abundance (Solar)\tablenotemark{b}  &  $1.0$ \nl
 &         Normalization\tablenotemark{c} &  $1.6 \times 10^{-5}$  \nl  
Power Law  &  Photon Index &  $1.7\pm 0.2$  \nl
      &         Normalization\tablenotemark{c} & $4.9 \times 10^{-5}$ \nl
\hline
 &  $\chi^2/\nu=71.2/60=1.2$  &  \nl

\enddata

\tablenotetext{a} {Absorption of CXB spectrum by Galactic disk has also been
taken into account. See \S 3.4.2 for details.}
\tablenotetext{b} {Frozen}
\tablenotetext{c} {Unabsorbed flux (${\rm photon \, s^{-1} \, cm^{-2} \,
keV^{-1}}$) at 10~keV}

\end{deluxetable}

\begin{deluxetable}{lll} 
\footnotesize 
\tablecaption{Best fit model parameters for the $3-35$~keV spectrum
from region R3 }
\tablewidth{470pt} 
\tablehead{ 
\colhead {Model Component} & 
\colhead {Parameter} & 
\colhead {Value} }   
\startdata 
 
Absorption\tablenotemark{a} &  $N_H$ ($10^{22}\,{\rm cm^{-2}}$) &  $0.1^{+0.6}_{-0.1}$
\nl 
Raymond-Smith & $kT$ (keV)  &  $3.9^{+0.9}_{-0.7}$  \nl 
 &        Abundance (Solar)\tablenotemark{b}  &  $1.0$ \nl 
 &         Normalization\tablenotemark{b} &  $2.3 \times 10^{-5}$  \nl 
Power Law  &  Photon Index &  $1.8^{+0.3}_{-0.1}$  \nl 
      &         Normalization\tablenotemark{c} & $5.0 \times 10^{-5}$ \nl
\hline 
 &  $\chi^2/\nu=80.4/60=1.3$  &  \nl

\enddata 
\tablenotetext{a} {Absorption of CXB spectrum by Galactic disk has also been
taken into account. See \S 3.4.2 for details.}
\tablenotetext{b} {Frozen} 
\tablenotetext{c} {Unabsorbed flux (${\rm photon \, s^{-1} \, cm^{-2} \, 
keV^{-1}}$) at 10~keV} 
 
\end{deluxetable}

\end{document}